\documentclass[acmsmall]{acmart}
\setcopyright{rightsretained}
\acmDOI{10.1145/3656426}
\acmYear{2024}
\copyrightyear{2024}
\acmSubmissionID{pldi24main-p300-p}
\acmJournal{PACMPL}
\acmVolume{8}
\acmNumber{PLDI}
\acmArticle{196}
\acmMonth{6}
\received{2023-11-16}
\received[accepted]{2024-03-31}
\AtBeginDocument{%
  \providecommand\BibTeX{{%
    \normalfont B\kern-0.5em{\scshape i\kern-0.25em b}\kern-0.8em\TeX}}}

\usepackage{amsmath,amsfonts}
\usepackage{algorithm}
\usepackage{algpseudocode}
\usepackage{graphicx}
\usepackage{textcomp}
\usepackage{xcolor}
\usepackage{xspace}
\usepackage{cleveref}
\usepackage{wrapfig}

\usepackage{pifont}
\usepackage{multirow}
\usepackage{multicol}
\usepackage{booktabs}

\usepackage{subcaption}
\usepackage{tabularx}
\usepackage{listings}
\usepackage{tikz}
\usetikzlibrary{shapes,arrows,calc}

\definecolor{keywordcolor1}{rgb}{0.5,0,0.5}
\definecolor{keywordcolor2}{RGB}{24, 23, 162}
\definecolor{textgray}{gray}{0.4}
\definecolor{mygray}{rgb}{0.5,0.5,0.5}

\usetikzlibrary{positioning,fit}

\def\BibTeX{{\rm B\kern-.05em{\sc i\kern-.025em b}\kern-.08em
    T\kern-.1667em\lower.7ex\hbox{E}\kern-.125emX}}

\newcommand{\newalg}{ISM\xspace}
\newcommand{\newalglong}{insert-sort-merge\xspace}

\newcommand*\colourcheck[1]{%
  \expandafter\newcommand\csname #1check\endcsname{\textcolor{#1}{\ding{52}}}%
}
\newcommand*\colourcross[1]{%
  \expandafter\newcommand\csname #1cross\endcsname{\textcolor{#1}{\ding{56}}}%
}
\newcommand*\colourslash[1]{%
  \expandafter\newcommand\csname #1slash\endcsname{\textcolor{#1}{\ding{52}\kern-1.1ex\raisebox{.7ex}{\rotatebox[origin=c]{125}{--}}}}%
}

\colourcheck{green}
\colourcross{red}
\colourslash{yellow}

\definecolor{codegreen}{rgb}{0,0.6,0}
\definecolor{codegray}{rgb}{0.5,0.5,0.5}
\definecolor{codepurple}{rgb}{0.58,0,0.82}

\lstdefinestyle{mystyle}{
    commentstyle=\color{codegreen},
    keywordstyle=\color{magenta},
    numberstyle=\tiny\color{codegray},
    stringstyle=\color{codepurple},
    basicstyle=\ttfamily\footnotesize,
    breakatwhitespace=false,         
    breaklines=true,                 
    captionpos=b,                    
    keepspaces=true,                 
    numbers=left,                    
    numbersep=5pt,                  
    showspaces=false,                
    showstringspaces=false,
    showtabs=false,                  
    tabsize=2
}

\lstdefinelanguage{lang}{
  keywords={split, reorder, fuse, pos, split, precompute},
  keywordstyle=\color{purple}\bfseries,
  identifierstyle=\color{black},
  sensitive=false,
  comment=[l]{//},
  alsoletter={=, +=},
}

\newcommand{\bnfdef}{\mathrel{::=}}
\newcommand{\bnfalt}{\mathrel{\mid}}
\newcommand{\mT}{\mathcal{T}}
\newcommand{\pluseq}{\mathrel{+}=}

\begin{document}

\title{Compilation of Modular and General Sparse Workspaces}

\author{Genghan Zhang}
\orcid{0000-0002-3866-8167}
\affiliation{%
  \institution{Stanford University}
  \country{USA}}
\email{zgh23@stanford.edu}

\author{Olivia Hsu}
\orcid{0000-0002-4195-8106}
\affiliation{%
  \institution{Stanford University}
  \country{USA}}
\email{owhsu@stanford.edu}
\orcid{}
\author{Fredrik Kjolstad}
\orcid{0000-0002-2267-903X}
\affiliation{%
  \institution{Stanford University}
  \country{USA}}
\email{kjolstad@stanford.edu}

\begin{abstract}
Recent years have seen considerable work on compiling sparse tensor algebra expressions. This paper addresses a shortcoming in that work, namely how to generate efficient code (in time and space) that scatters values into a sparse result tensor. We address this shortcoming through a compiler design that generates code that uses sparse intermediate tensors (sparse workspaces) as efficient adapters between compute code that scatters and result tensors that do not support random insertion.
Our compiler automatically detects sparse scattering behavior in tensor expressions and inserts necessary intermediate workspace tensors. We present an algorithm template for workspace insertion that is the backbone of our code generation algorithm. Our algorithm template is modular by design, supporting sparse workspaces that span multiple user-defined implementations.
Our evaluation shows that sparse workspaces can be up to 27.12$\times$ faster than the dense workspaces of prior work. On the other hand, dense workspaces can be up to 7.58$\times$ faster than the sparse workspaces generated by our compiler in other situations, which motivates our compiler design that supports both. Our compiler produces sequential code that is competitive with hand-optimized linear and tensor algebra libraries on the expressions they support, but that generalizes to any other expression. Sparse workspaces are also more memory efficient than dense workspaces as they compress away zeros. This compression can asymptotically decrease memory usage, enabling tensor computations on data that would otherwise run out of memory.
\end{abstract}

\ccsdesc[500]{Software and its engineering~Domain specific languages}
\ccsdesc[500]{Software and its engineering~Source code generation}


\maketitle

\section{Introduction}
\label{sec:introduction}
Sparse tensor algebra is an important class of computation used in various applications~\cite{kipf2017gcn,liu2015sparse,kolda2008decomposition,Bell2012ESC,kepner2011graph,kolda2009tensor}. It generalizes linear algebra to higher-order tensors, where the tensors may be dense or sparse. 
Domain-specific sparse tensor algebra compilers~\cite{kjolstad:2017:taco,strout2018spf,zheng2022sparta,ye2023sparsetir} automatically generate and optimize sparse tensor algebra code. These compilers are becoming more prevalent because they can generate codes for the large combination of tensor algebra expressions, compressed data structures, optimizations, and hardware backends that are not supported by libraries~\cite{henry_hsu:2021:array,yadav:2022:distal,zhang2023sgap,won2023unified}. 

\begin{figure}
\centerline{\includegraphics[width=0.99\textwidth]{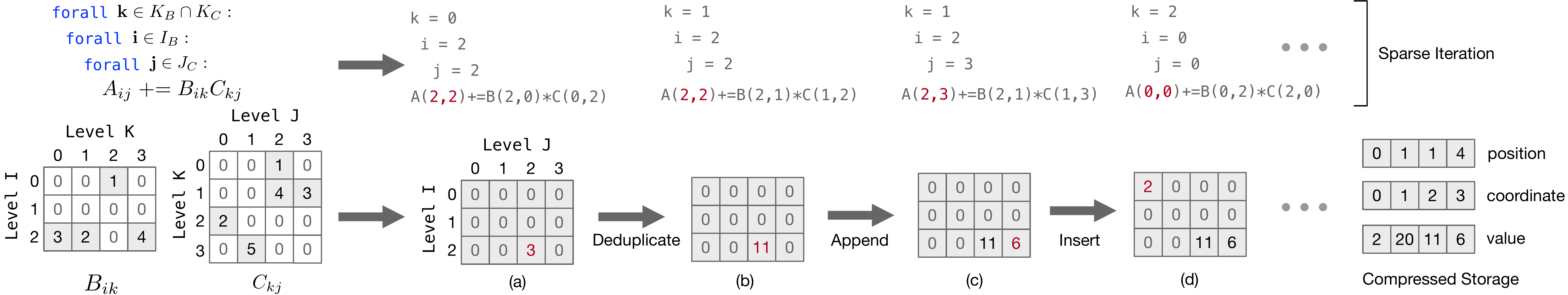}}
\caption{A second-order dense workspace for outer-product matrix multiplication (SpGEMM). The above for-loop pseudo codes show the sparse iterations that generate tensor components. Red numbers represent newly generated coordinates and values. The workspace must support three behaviors: deduplicating (a $\rightarrow$ b), appending (b $\rightarrow$ c), and inserting (c $\rightarrow$ d). The computation utilizes a workspace since the final compressed data structures do not support insertion. Furthermore, the result storage should be compressed for memory efficiency since the final output has only four values.} 
\label{fig:spwsintro}
\end{figure}

However, there is a hole in the above sparse compiler work: the sparse scattering problem. Sparse scattering happens when a sparse result tensor is written to in an arbitrary order. This is a common problem in sparse tensor algebra~\cite{liu1992solver,pal2018outerspace,yang2023ISOSceles}. \Cref{fig:spwsintro} shows a concrete example of sparse scattering. 
In this example, the tensor component needs to be inserted in front of components that have already been placed. 
Specifically, the tensor component generated in state (d) has coordinates that are lexicographically smaller than the coordinates from states (a)--(c). However, the result matrix is stored in a compressed data structure, which does not permit efficient insertion into this location. Therefore, a temporary tensor is necessary as an adapter between the computed matrix components and the result tensor storage. Such a temporary is called a workspace. 

\begin{figure*}
\centerline{\includegraphics[width=\linewidth]{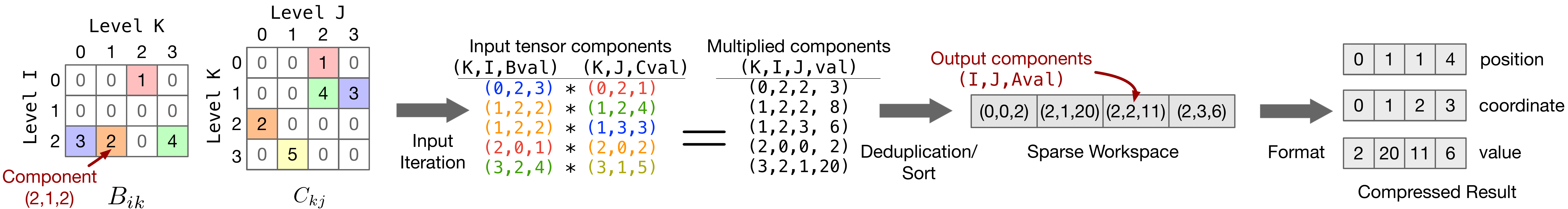}}
\caption{A second-order sparse workspace for the outer-product SpGEMM in \Cref{fig:spwsintro}. The colored nonzero components of the input tensors show a correspondence to their respective input tensor components. In a second-order workspace, each \texttt{(I,J,val)} tensor component is indexed by two variables \texttt{I} and \texttt{J}.}
\label{fig:sp-ws-desc}
\end{figure*}

\Cref{fig:spwsintro} and \Cref{fig:sp-ws-desc} show examples of a dense and a sparse workspace, respectively. A dense workspace holds the temporary values in a dense array whereas a sparse workspace stores temporary values in compressed data structures. In order to store a tensor component (both coordinates and the value) into any workspace, the value must first be calculated from the inputs. The value is then summed with the existing value in the workspace during the deduplication step. Then, the deduplicated value is either appended to the end of or inserted into the middle of the workspace data structure. An example of this process is shown in \Cref{fig:spwsintro}. In the case of a sparse workspace (as in \Cref{fig:sp-ws-desc}), it is common to only append values to the compressed data structure (without insertion) for efficiency. In order to ensure that the values end up in the correct output order, sparse workspace algorithms include a sorting step.

A sparse workspace is essential when sparse scattering into high-order tensors because dense higher-order storage would require too much memory. Sparse tensors are often asymptotically sparse, which empirically implies that less than $1\%$ of values are nonzero~\cite{kjolstad:2020:phd-thesis}. For example, multiplying the ``marine1'' matrix (from chemical oceanography~\cite{kolodziej2019suitesparse}) by itself transposed would require a dense workspace that is $5132\times$ larger than the sparse workspace. Beyond SpGEMM, it is common to have higher-order workspaces when the input and output tensors are higher-order, as in the sparse convolution~\cite{yang2023ISOSceles}.

We would like to support sparse workspaces in sparse compilers, but there are two challenges: 
\begin{enumerate}
    \item The compiler should be modular to allow users to automatically plug in and combine various optimization strategies. Workspace policies are diverse~\cite{demouth2012cusparse,Buluc2009heap,Bell2012ESC}, so modularity is essential for generating code that is competitive with libraries.
    \item The compiler should be sufficiently general so as to produce code for any expressions with sparse scattering. That means the sparse workspaces generated by the compiler must handle tensors of any order and with any compression format.
\end{enumerate}

To address these challenges, we propose an algorithm template for sparse workspaces and a compiler that integrates hand-written implementations of the template functions to generate correct and efficient code. Generating code by combining templated code with filled-in, user-written functions (also called codelets or stencils) has been used throughout history.
This idea of \textit{code composition} through templates is widely used in compilers for domain-specific languages (DSLs)~\cite{stichnoth1997composition}. Specific examples include compiling relational algebra queries~\cite{neumann2011efficiently},FFTs~\cite{frigo1998fftw}, and more recently tensor algebra~\cite{bansal2023mosaic,exocompilation} and neural networks~\cite{chen2018tvm,ansel2024pytorch}. Beyond DSLs, general programming systems have also benefited from similar compositional methods through techniques like template JIT~\cite{vcode,ertl2004jit,xu2021copy,pichler2023hybrid}. We are, however, the first to propose a modular template-based approach for compiling sparse tensor algebra expressions with scattering behavior. 

For the algorithm template, we design a modular representation of sparse workspaces expressive enough to describe several prior, efficient sparse workspace policies. For the compiler, we incorporate our representation to the TACO system~\cite{kjolstad:2017:taco} to provide general support for generating sequential implementations of sparse tensor algebra expressions with sparse workspaces. We design user interfaces at different levels of our system to express new policies. Our contributions are:
\begin{itemize}
\item An analysis framework that categorizes sparse tensor algebra expressions with respect to how they assemble the result.
\item An algorithm template for sparse workspace generation. The algorithm generalizes to workspaces implemented with various compressed data structures and optimization policies. 
\item An automatic workspace insertion algorithm that transforms expressions to include those workspaces that are necessary for correctness.
\item Extensions to the TACO programming model and intermediate representation to generate code for expressions with sparse workspaces. 
\end{itemize}

We evaluate these contributions by comparing them against current dense workspace techniques. Our evaluation shows that sparse and dense workspaces do not dominate each other but are useful in different situations, motivating the need for both approaches. Specifically, dense workspaces can perform up to 7.58$\times$ faster than sparse workspaces, whereas sparse workspaces can perform up to 27.12$\times$ faster than dense ones, depending on the input data. Our compiler can generate general tensor algebra codes with both sparse and dense workspaces while still producing code competitive with hand-optimized linear and tensor algebra libraries. Furthermore, our compiler produces code with a 3.6$\times$ improvement in memory footprint on average (geomean) when compared to dense workspaces that fit in our machine's memory. The sparse workspace code generated by our compiler is not limited by the machine's memory for any input data, but the dense workspace is unable to fit in memory for 10 out of the 60 input matrices run. Therefore, our compiler scales better both in performance and memory footprint for higher-order and larger data sizes. Though our approach generates sequential sparse workspace code, it provides a foundation for developing a code generator that also supports parallel sparse workspace codes. 

\section{Sparse tensor algebra expression taxonomy}
\label{sec:taxonomy}
We introduce an analysis framework to identify when sparse scattering occurs in sparse tensor algebra expressions. Sparse scattering
can be classified based on the relationship between the tensor access order and the expression loop order. These orderings, defined in \Cref{sec:idxorders}, determine how input tensors are accessed and how the result tensor is assembled. 
Based on these orders, we classify sparse tensor algebra expressions along two axes (\Cref{sec:classification}) and place our work in the context of prior work by showing how they lack the ability to handle sparse scattering (\Cref{sec:thetable}).

\subsection{Index Variable Orderings}
\label{sec:idxorders}
\begin{figure}

\begin{minipage}{0.45\linewidth}
\footnotesize
\centering
\[
\begin{array}{rlrlrl}
\textit{Index Variable} & i & \textit{Constant} & c & \textit{Tensor} & \mT \\
\end{array}
\]
\vspace*{-1.5em}
\[
\begin{array}{rlcl}
  \textit{Access} & a & \bnfdef & \mT_{i*} \\
  \textit{Expression} & e & \bnfdef & a \bnfalt c \bnfalt e + e \bnfalt e \cdot e \bnfalt \ldots \\
  \textit{Statement} & S & \bnfdef & \forall_{i*}~S \bnfalt a = e \bnfalt a \pluseq e \bnfalt \ldots
  \\ 
\end{array}
\]
\caption{A simplified concrete index notation (CIN) syntax with no scheduling relationships. }
\label{fig:expression-syntax}
\end{minipage}%
\hfill
\begin{minipage}{0.52\linewidth}
    \centerline{\includegraphics[width=\textwidth]{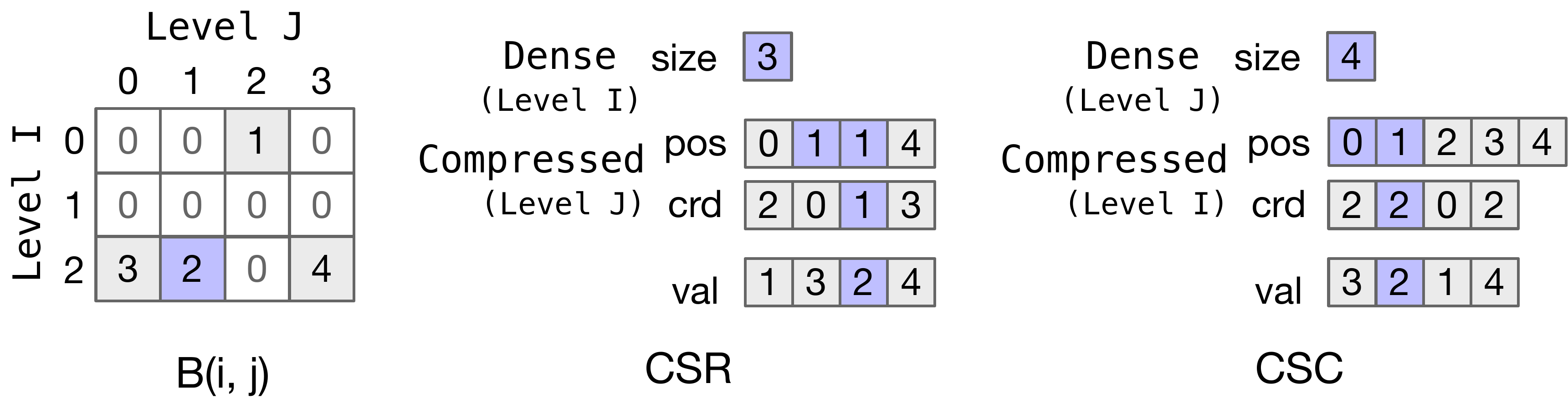}}
    \caption{Two example tensor level formats for compressed sparse row (CSR) and compressed sparse column (CSC).}
    \label{fig:level-example}
\end{minipage}

\end{figure}

Sparse tensor algebra expressions, given in concrete index notation (CIN)~\cite{kjolstad:2019:workspaces}, are composed of tensors, index variables, and forall nodes.
CIN is a loop-based intermediate representation in which physical tensor representations are abstracted away. Below the tensor abstraction, tensors are stored level by level. Each level describes the coordinates of one tensor mode and can be materialized as a data structure in some format.  \Cref{fig:expression-syntax} gives the syntax of core parts of the CIN this paper uses.

Sparse tensor algebra expressions along with their tensor formats have two types of ordering properties: access orders and loop orders. An access order describes how an individual tensor \textit{needs} to be accessed (read) and assembled (written to). The loop order, on the other hand, determines how each tensor in an expression \textit{is actually} accessed and assembled. The loop order is directly defined as the order of loop index variables---index variables next to $\forall$ nodes in \Cref{fig:expression-syntax}. 

The access order defines the order of the index variables that access the physical storage of a tensor. It is generated by an \texttt{AccessMap} function that maps the index variables used to index (access) a tensor to the storage levels of its format. The access index variables are the index variables inside an Access in \Cref{fig:expression-syntax}. 
The $AccessMap$ builds upon the established level formats such as $Dense$ and $Compressed$ in \citet{kjolstad:2017:taco} 
and the coordinate mapping described in~\citet{chou:2020:conversion}.

Let us consider outer-product SpGEMM $A_{ij} = \sum_k B_{ik} C_{kj}$ where $A$ and $C$ are stored in the compressed sparse rows format (CSR) and $B$ is stored in compressed sparse columns (CSC) (shown in \Cref{fig:level-example}). In CIN, this is expressed as $\forall_{kij} \; A_{ij} \pluseq B_{ik} C_{kj}$ with the following tensor AccessMaps:
\begin{enumerate}
\footnotesize
    \item $\text{AccessMap}_A(\{i,j\},\{Dense, Compressed\})=\{i,j\}$,
    \item $\text{AccessMap}_B(\{i,k\},\{Dense, Compressed\})=\{k,i\}$, and
    \item $\text{AccessMap}_C(\{k,j\},\{Dense, Compressed\})=\{k,j\}$.
\end{enumerate} Therefore, the access orders for $A$, $B$ and $C$ are $i\rightarrow j$, $k\rightarrow i$, and $k\rightarrow j$ respectively, and the loop order for the expression is $k\rightarrow i\rightarrow j$. 

\subsection{Classification}
\label{sec:classification}

\begin{figure}
\centerline{\includegraphics[width=0.99\textwidth]{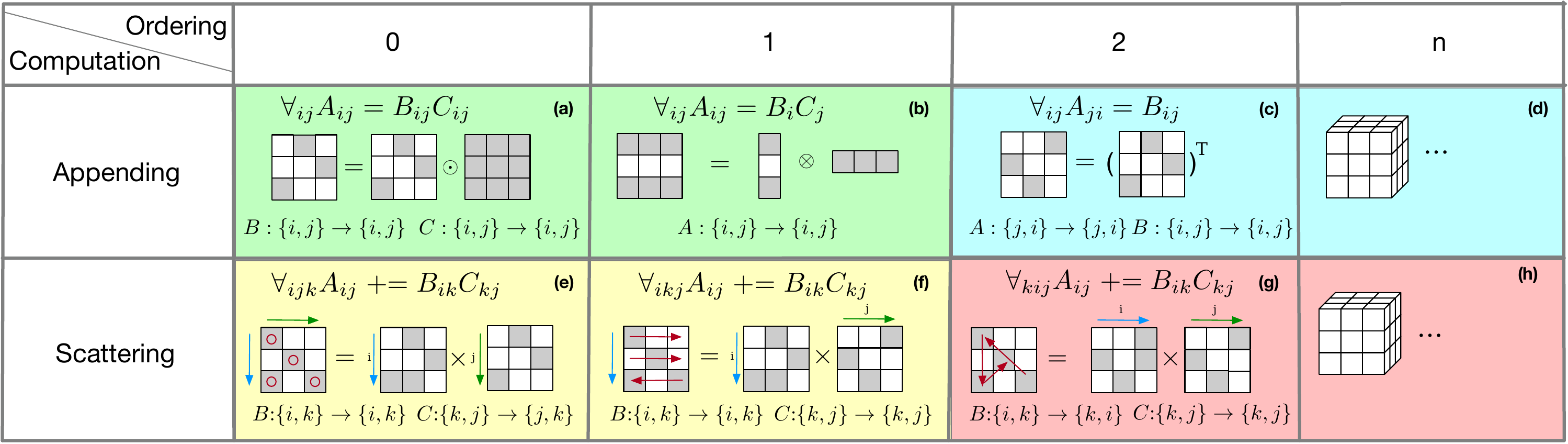}}
\caption{Sparse tensor algebra expressions classified by computation and ordering. Blue and green arrows show the loop order and red lines show the result assembly order. Tensors' index variables encode access order. } 
\label{fig:sp-scatter}
\end{figure}

We classify sparse tensor algebra expressions based on the computation and ordering of the result coordinates as shown in \Cref{fig:sp-scatter}. The computation axis describes whether each resulting value corresponds to a single value computed from the operands or whether it corresponds to a combination of many computed values (e.g., a sum).
The ordering axis describes 
how the result coordinates are generated, and to what extent the generation order matches the access order of the output tensor. Intuitively, the order of input sparse iteration with respect to the order of output access decides the number of dimensions we need in the intermediate workspace data structure.
For example, an inner product matrix multiplication requires only a scalar (order 0) temporary, while an outer matrix multiplication requires a matrix (order 2) temporary. 

The result coordinate calculation is \textit{appending} if the pattern of nonzeros of the result is the same as the input iteration space and \textit{scattering} otherwise. As shown in \Cref{fig:sp-scatter}(a), the element-wise multiplication of a sparse and dense matrix is appending because the sparse result matrix $A$ has the same coordinates as the sparse input matrix $B$. Tensor transpositions like \Cref{fig:sp-scatter}(c) are also appending because the input coordinates do not require an intersection or union to compute the result coordinates even though they are transposed.

The result coordinates likely needs to be assembled in a \textit{sparse} way if the result coordinate ordering can not be narrowed to a first-order (vector) or zero-order (scalar) tensor as the loop proceeds, due to the large worst-case memory cost of storing a sparse matrix or tensor in a dense data structure. In other words, the assembly is sparse if the loop order mismatches with the output access order at a position greater than one. As shown in \Cref{fig:sp-scatter}(f), the loop order of the row-wise SpGEMM is $i\rightarrow k\rightarrow j$, and the output access order is $i\rightarrow j$. These two orders only mismatch on index $k$ with $k$ at the first position (from the inner-most index), so we classify the expression as a first-order \textit{dense} scattering (the yellow area in \Cref{fig:sp-scatter}). On the contrary, the loop order of the outer-product SpGEMM in \Cref{fig:sp-scatter}(g) is $k\rightarrow i\rightarrow j$, which mismatches with the output access order with $k$ at the second position. Therefore, we classify the expression as second-order \textit{sparse} scattering (the red area in \Cref{fig:sp-scatter}). 

In general, ordering is determined by the position of the first index variable in the loop order that does not match the access order. Given a loop order $\mathcal{L} = i_M \rightarrow i_{M-1} ... \rightarrow i_1$, and the output tensor's access order $\mathcal{A} = j_N \rightarrow j_{N-1} ... \rightarrow j_{1}$, 
there are two important positions where the orders mismatch. The first position occurs during tensor transpositions, where $p_1$ is the position of the first index variable from the left in $\mathcal{A}$ that mismatches with $\mathcal{L}$. The second is $p_2$, the position of the first non-access index variable $i$ in $\mathcal{L}$. If $i$ is before the first access index variable $j_1$, then $p_2=0$ because output tensor components are accumulated scalar by scalar. If $i$ is after the last access index variable $j
_N$, then $p_2=N$  because all components in the output tensor are accessed multiple times. Otherwise, the $p_2$ equals the $i$'s position in the middle of $\mathcal{L}$ between $j_N$ and $j_1$. The result coorddinates' ordering equals $max(N+1-p_1, p_2)$. The materialized workspace cannot have fewer orders than the ordering. We provide a more detailed algorithm that determines the required order of a workspace in \Cref{sec:compiler-auto-insert}.

\subsection{Limitation of Prior Work}
\label{sec:thetable}
As shown in~\Cref{tab:comparison}, prior sparse compilers~\cite{kjolstad:2017:taco,kjolstad:2019:workspaces,zhao2023spf,ye2023sparsetir} cannot generate code for expressions with sparse scattering behavior. 
Although compilers 
can generate efficient co-iteration on sparse input tensors~\cite{strout2018spf,zheng2022sparta}, they often assume the output compression format is known beforehand as dense or identical to one of the compressed inputs~\cite{zhao2023spf,ye2023sparsetir}. Compilers that generate code with dense workspaces~\cite{bik1994automatic,bik:2022:mlirsparse,william1998sipr,kjolstad:2019:workspaces,scott2023index} can assign a temporary dense array to hold values generated from the input co-iteration. However, they either only operate on linear algebra~\cite{bik1994automatic} or consume too much memory when densifying a sparse higher-order tensor into the temporary~\cite{kjolstad:2019:workspaces}. The TACO body of work provides format conversions~\cite{chou:2020:conversion} that can handle assignment expressions where the result tensor has the same elements as the input but with varying formats. However, it cannot process dynamic, out-of-order nonzeros generated by sparse iteration with scattering. 
As such, our work is the first sparse tensor algebra compiler that takes in dynamic components from input tensors and scatters them into tensors of any format.

\begin{table*}
\caption{Output tensor support in prior work. The \greencheck~ denotes full support and \redcross~ denotes no support.}
\scriptsize
\centering
\begin{tabular}{lccccc}
\toprule
    \multicolumn{1}{c}{\multirow{4}{*}{Sparse Tensor Algebra Compilers}} & \multicolumn{2}{c}{Output Appending} & \multicolumn{3}{c}{Output Scattering} \\
    \cmidrule(lr){2-3} 
    \cmidrule(lr){4-6}
     & \multirow{2}{*}{Dense} & \multirow{2}{*}{Sparse} & \multirow{2}{*}{Dense} & \multicolumn{2}{c}{Sparse} \\ 
     \cmidrule(lr){5-6}
     & & & & Dense workspace & Sparse workspace \\
    \midrule
    SparseTIR~\cite{ye2023sparsetir} & \greencheck & \redcross & \redcross & \redcross & \redcross \\
    Sparse Polyhedral Framework~\cite{zhao2023spf} & \greencheck & \redcross & \redcross & \redcross & \redcross \\
    MLIR Sparse Dialect~\cite{bik:2022:mlirsparse} & \greencheck & \redcross & \greencheck & \greencheck & \redcross \\
    TACO with dense workspaces~\cite{kjolstad:2019:workspaces} & \greencheck & \redcross & \greencheck & \greencheck & \redcross \\
    TACO with format conversion~\cite{chou:2020:conversion} & \greencheck & \greencheck & \redcross & \redcross & \redcross \\
    Our work & \greencheck & \greencheck & \greencheck & \greencheck & \greencheck \\
    \bottomrule
\end{tabular}
\label{tab:comparison}
\end{table*}
\section{Overview}

\tikzstyle{newblock} = [rectangle, draw=none, fill=blue!20, text centered, minimum height=3em, scale=0.70]
\tikzstyle{block} = [rectangle, draw, text width=5em, text centered, minimum height=3em, scale=0.70]
\tikzstyle{input} = [text centered, align=left, draw=none, fill=none, scale=0.70]
\tikzstyle{output} = [coordinate]
\begin{figure}[b]
    \centering
    \begin{tikzpicture}[auto, font=\scriptsize, scale=0.70, node distance=3cm,>=latex']

    \node [block] (cin) {Concrete Index Notation (CIN)};
    \node [block, right=2cm of cin] (llir) {Low-Level \\ Imperative IR};

    \node [newblock, text width=5em, right=1.2cm of llir] (ism) {Insert-Sort-Merge Calls \\ (\Cref{sec:compiler-code-gen})};
    \node [fit=(ism), draw, rectangle, inner sep=10pt, label={[font={\bfseries\scriptsize}, below]Imperative Code}] (code) {};

    \node [newblock, text width=8em, above=1.2cm of ism] (policy) {Sparse Workspace Policies \\ (\Cref{sec:policy})};
    
    \draw [->] (policy) -- node {} (code);
    \draw [->] (llir) -- node {} (code);
    \draw [->] (cin) -- node[name=lowerer, font={\bfseries\footnotesize}] {Lowerer} (llir);
    
    \coordinate (lowererN) at (lowerer.north);
    
    \node [newblock, text width=1.5cm, above=0pt of lowererN] (insertsortmerge) {Insert-Sort-Merge \\ Algorithm Template \\ (\Cref{sec:alg-template})};

    \node [fit=(cin)(llir)(insertsortmerge), draw, rectangle, inner sep=5pt, label=below:Tensor Algebra Compiler] (compiler) {};

    \node [input, left=1cm of compiler, font={\bfseries\scriptsize}] (algnode) {Tensor Index Notation};
    \coordinate (rightinput) at (algnode.east);
    
    \node [newblock, text width=7em, above=0.5cm of algnode] (userfn) {User-Defined \\ Template Functions \\ (\Cref{sec:compiler-sched})};
    \node [fit=(userfn), draw, rectangle, inner sep=10pt, label={[font={\bfseries\scriptsize}, below]Format Language}] (formatnode) {};
    
    \node [newblock, text width=7em, below=0.5cm of algnode] (mansp) {Sparse Workspace Insertion \\ (\Cref{sec:compiler-sched})};
    \node [newblock, text width=8em, left=1em of mansp] (autosp) {Automatic Workspace Reasoning \\ (\Cref{sec:compiler-auto-insert})};
    \node [fit=(mansp)(autosp), draw, rectangle, inner sep=10pt, font={\bfseries\scriptsize},label={[font={\bfseries\scriptsize}, below]Scheduling Language}] (schednode) {};
    
    \draw [->] (formatnode) -- node {} (compiler);
    \draw [->] (schednode) -- node {} (compiler);
    \draw [->] (algnode) -- node {} (compiler);
    \coordinate (policyleft) at (policy.north west);
    \coordinate (formatright) at (formatnode.north east);
    \coordinate (policyShifted) at ($(policyleft) + (0, -0.40cm)$);
    \coordinate (formatShifted) at ($(formatright) + (0, -0.4cm)$);
    
    \draw [->] (policyShifted) -- node {} (formatShifted);
    \end{tikzpicture}
    \caption{System Overview. Blue components denote new contributions of this work.}
    \label{fig:overview}
\end{figure}
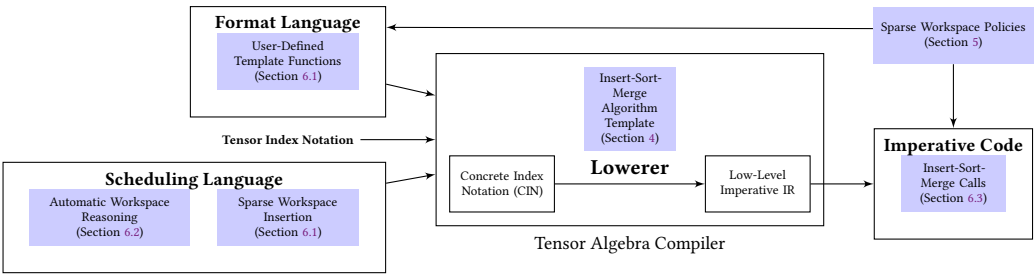

We implemented the new compiler techniques for sparse workspaces as an extension to the open-source tensor compiler TACO~\cite{kjolstad:2017:taco}, but these technique can also be used in other sparse tensor compilers~\cite{ye2023sparsetir,bik:2022:mlirsparse,mutlu2020comet}. \Cref{fig:overview} gives an overview of our new compiler with sparse workspaces. It takes as input an expression in tensor index notation (also called Einsum), a format language~\cite{chou2018format} that encodes the AccessMap, and a scheduling language~\cite{senanayake:2020:scheduling} that incorporates sparse workspaces. Our compiler combines these three input languages into the CIN intermediate representation. 

We propose the \newalglong (\newalg) algorithmic template for generating sparse workspace code that can be inserted into generated sparse tensor algebra computation loop nests. The \newalg algorithm is the algorithmic backbone that gives us a straightforward way to generate modular sparse workspace code (\Cref{sec:alg-template}). Whenever a CIN expression contains a sparse workspace tensor, our compiler lowers it into dense and sparse loops with the \newalg template inserted. The \newalg template only defines abstract memory pools and function interfaces, with holes for different sparse workspace building blocks. During code generation, our compiler fills in those holes with user-defined workspace implementations (\Cref{sec:compiler-code-gen}).

We explore several concrete sparse workspace policies in \Cref{sec:policy} that are compatible with the \newalg template. These are example user-defined workspace policies that materialize the \newalg template during code generation, but many more are possible. Users provide their \newalg template function implementations as input into the format language of our compiler (\Cref{sec:compiler-sched}).

Our compiler technique operates across various stages of compilation and bridges prior work on sparse iteration~\cite{kjolstad:2017:taco,henry_hsu:2021:array} with prior work on format conversions~\cite{chou:2020:conversion} using the concepts from the \newalg template.
Apart from the format language, we extend the scheduling language to include sparse workspaces. 
For user productivity, we also automate the insertion of sparse workspaces to ensure code correctness using a compiler transformation (\Cref{sec:compiler-auto-insert}). Then, we extend the lowerer from CIN to C++ to include the \newalg memory pools and function interfaces (\Cref{sec:compiler-code-gen}).

\section{Algorithm Template for Sparse Workspaces}
\label{sec:alg-template}

The \newalglong algorithm template (\newalg) is a four-stage algorithmic backbone for constructing sparse workspaces. It describes the mechanism we use to insert sparse workspaces into any sparse tensor algebra expression. The compiler emits codes that materialize the template, as described in \Cref{sec:compiler}. 

\newalg accumulates the input tensor components, stores them to a temporary array, and finally converts the temporary array to the result's data structure. We leave the concrete memory data structure decisions and function definitions up to the user, allowing for a wide variety of concrete sparse workspace policies as described in \Cref{sec:policy}.

\subsection{Tensor Component Abstraction}
\begin{figure*}
\centerline{\includegraphics[width=\linewidth]{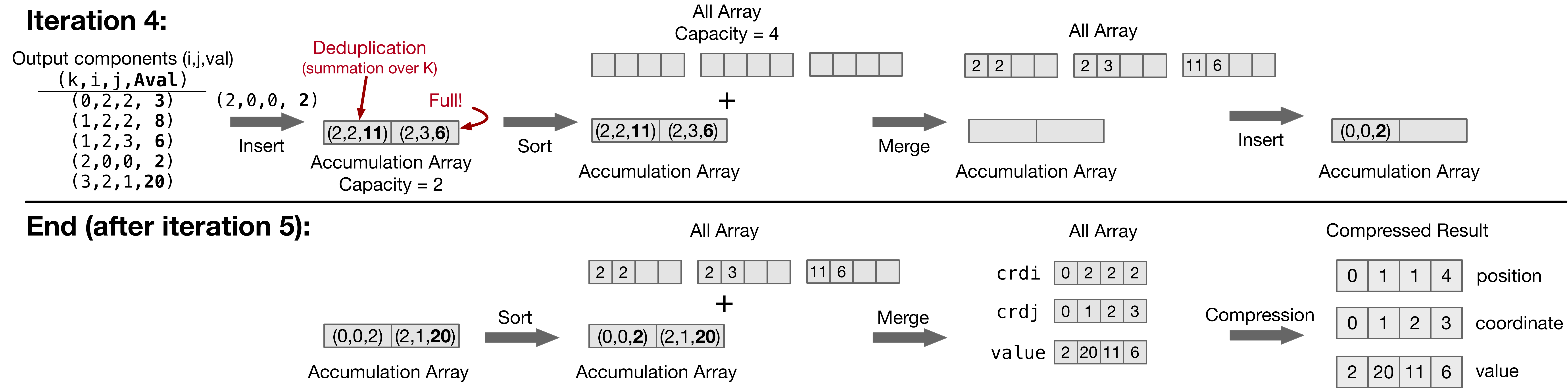}}
\caption{The \newalglong algorithm template on our outer-product SpGEMM example from \Cref{fig:sp-ws-desc}.}
\label{fig:ism}
\end{figure*}

We refer to a nonzero value and its coordinates as a tensor component. The input to the \newalg algorithm is a stream of tensor components to be accumulated into the result tensor.
As shown in \Cref{fig:ism}, an output tensor component is composed of coordinates \texttt{i} and \texttt{j}, and the value \texttt{val}, which fully describes an element in a matrix. Although tensor components are less memory efficient per nonzero element than other compressed representations such as CSR, they are a direct and clean abstraction that simplifies the algorithm template and code generation.

\subsection{The Accumulation Array and All Array}
\label{sec:two-array}
\newalg requires storage in which to accumulate components, deduplicate components with the same coordinates, and compress all the generated components to the result. We achieve this by defining two abstract memory pools: the \textit{accumulation array} and the \textit{all array}.
The all array is required since a sparse workspace must include at least one memory data structure that stores all of the input components to assemble the output result tensor. To improve performance and manage duplicates, the accumulation array acts as a landing pad that batches the insertions to the all array.

The all array is a temporary linear storage for the result components. Generated components are scattered into it in sorted order, and then the result components are extracted from it to assemble the final tensor format.
As shown in \Cref{fig:ism}, the all array stores coordinates in level \texttt{I} and \texttt{J}, and the values for each unique result tensor component. Finally, the all array is compressed to the result data structure, for example, the CSR format in \Cref{fig:ism}.

We design an accumulation array to serve as an intermediate buffer between the generated tensor components and the all array. The accumulation array can be materialized as any efficient data structure on higher-order tensor components that supports random insertion. In \Cref{fig:ism}, the accumulation array deduplicates at (2,2) by adding the values of components (2,2,3) and (2,2,8). When the accumulation array is full, it is merged with the all array, which we introduce in \Cref{sec:fourstage}.

\begin{wrapfigure}{r}{0.47\linewidth}
\vspace{-2em}
\begin{minipage}{\linewidth}
\footnotesize
\begin{algorithm}[H]
\caption{The \newalg algorithm template}\label{alg:ISM}
\begin{algorithmic}[1]
\State \textbf{inputs:} Value arrays $val_t$ of tensor $t$, Accumulation array $Acc$, All array $All$. 
\State \textbf{output:} Final output tensor $Out$
\State 
\State $Allocate(Acc)$
\While{there's still nonzero}
    \State Iterate and append coordinate to $crds$
    \If{reach the last level}
        \State $Insert(crds,Expression(\{Val_t\}),Acc)$
        \If{$Acc.full$}
            \State $Sort(Acc)$
            \State $Merge(Acc,All)$
            \State $Insert(crds,Expression(\{Val_t\}),Acc)$
        \EndIf
    \EndIf
\EndWhile
\If{not $Acc.empty$}
    \State $Sort(Acc)$
    \State $Merge(Acc,All)$
\EndIf
\State $Compress(All,Out)$
\end{algorithmic}
\end{algorithm}
\end{minipage}
\vspace{-2.5em}
\end{wrapfigure}

The accumulation array improves performance since it divides and conquers the insertion of components into the all array. We provide evidence of this benefit through a Big-O analysis in \Cref{sec:bigo-appendix} and empirically in \Cref{sec:eval-design-choices}.
With the accumulation and all arrays defined, we will describe how they interact in the \newalglong algorithm template to create the result tensor.

\subsection{Four-stage Template Model}
\label{sec:fourstage}
The input tensor components---generated by the loops that iterate over and compute on sparse and dense tensors---are processed through the \newalglong algorithm and stored into the accumulation and all arrays. Any sparse workspace algorithm must support two types of computation: insertions and deduplications (\Cref{fig:spwsintro}). 
Insertions place generated tensor components into memory and deduplications sum inserted components that have the same coordinates (collisions). 
The design of the \newalglong algorithm template, shown in \Cref{alg:ISM}, distills the sparse workspace construction process into four stages:
\begin{enumerate}
    \item \textbf{Insertion.} The insertion stage inserts tensor components into the accumulation array.
    \item \textbf{Sorting.} The sorting stage triggers when the accumulation array fills up and sorts its components into the order of the result tensor storage. 
    \item \textbf{Merging.} The merging stage merges the components from the accumulation array into the all array and clears the accumulation array.
    \item \textbf{Compression.} The compression stage transforms components stored as coordinates in the all array to the result data format.
\end{enumerate}

In the \newalg algorithm, generated components are inserted into the accumulation array during the insertion stage until the array's capacity is reached (\Cref{alg:ISM} line 8). Tensor components with the same coordinates are reduced (summed) either during the insertion stage or the sorting stage. When the accumulation array is full, the ISM enters the sorting stage (\Cref{alg:ISM} line 9). 

Sorting of the accumulation array and merging of the accumulation array into the all array occurs before the accumulation array is cleared and a new tensor component is inserted. The sorting algorithm implementation depends on the choice of accumulation array data structure. We describe multiple concrete sorting and data structure implementations in \Cref{sec:policy} along with their tradeoffs. Next, the merging stage moves components from the accumulation array into the all array, where components remain sorted, and components with equivalent coordinates are reduced. The \newalg algorithm separates the sorting phase from the merging phase because sorting can reduce the complexity of merging from $O(n\times m)$ to $O(n+m)$ where $n$ and $m$ denote the number of components in the all and accumulation arrays respectively. When the accumulation array is cleared, the \newalg algorithm enters a new iteration.

After the last nonzero element is processed, the remaining components in the accumulation array are sorted and merged into the all array (\Cref{alg:ISM} lines 16--18). The last step of the \newalglong algorithm template is to compress the final all array to the expected output tensor format (\Cref{alg:ISM} line 20). \Cref{fig:spgemm-code} shows the imperative code that materialize each part of \Cref{alg:ISM} for outer-product SpGEMM example in \Cref{sec:idxorders}.

\begin{figure} 
    \scriptsize
    \begin{minipage}[b]{\linewidth}
    \centering
    \begin{lstlisting}[xleftmargin=2em,language=C++,commentstyle=\color{gray},multicols=2,numbers=left,framexleftmargin=3em]
#include "ism.h"
// Allocate
AccArray Acc(2,cap,"Coord");
AllArray All(2,cap);
Component c;

// Insert-Sort-Merge
for (int k = 0; k < K; k++) {
  for (int iB = B.pos[k]; iB < B.pos[k+1]; iB++) {
    int i = B.crd[iB];
    c.crd[0] = i;
    for (int jC = C.pos[k]; jC < C.pos[k+1]; jC++) {
      int j = C.crd[jC];
      c.crd[1] = j;
      c.val = B.val[iB] * C.val[jC];
      Insert(c, &Acc);
      if (Acc.full) {
        All.realloc(Acc.size);
        Sort(&Acc);
        Merge(&Acc, &All);
        Acc.refresh();
        Insert(c, &Acc);
      }
    }
  }
}
if (Acc.size > 0) {
  All.realloc(Acc.size);
  Sort(&Acc);
  Merge(&Acc, &All);
}

// Compress
A.crd = All.crd[1];
A.val = All.val;
int* A.pos = (int*)calloc(I+1, sizeof(int));
int iw = 0;
while (iw < All.size) {
  int i = All.crd[0][iw];
  int segend = iw + 1;
  while (segend < All.size && 
          All.crd[0][segend] == i) {
    segend++;
  }
  A.pos[i + 1] = segend - iw;
  iw = segend;
}
int cnt = 0;
for (int pA = 1; pA < I + 1; pA++) {
  cnt += A.pos[pA];
  A.pos[pA] = cnt;
}
    \end{lstlisting}
    \caption{\label{fig:spgemm-code}
        Simplified C++ code generated for outer-product SpGEMM following the \newalg template. The header file contains definitions for the accumulation array, the all array, and the \newalg functions. The compression stage transforms \texttt{All} from COO to the result format CSR.
    }
    \vspace{4ex}
  \end{minipage}
\end{figure}

\subsection{Recreating Prior Work with the \newalg Framework}
\label{sec:prior-ism}

The \newalglong algorithm template provides a general and modular framework to assemble various concrete implementations. We will show how existing workspace algorithms from prior work are expressed in terms of the \newalg template. 
As shown in \Cref{tab:kernel-ism}, prevailing hand-optimized workspace algorithms for SpGEMM can be expressed in \newalg by implementing different sorting and merging algorithms. Since these algorithms do not use an accumulation array, the merge stage column in ~\Cref{tab:kernel-ism} shows how their workspaces are compressed to the output. 

For example, Gustavson's algorithm uses a flag array to label positions that already have nonzero elements. Therefore, it can deduplicate components upon insertion by checking the flags. Meanwhile, the components are sorted because the position in the 1-D coordinate list is equal to the level J index, which is based on a bucket sort~\cite{devroye1986bucketsort}. The positions where the workspace flag is set are merged to the output; this procedure is also known as boolean indexing~\cite{cleveland1984boolindex}. Other algorithms can be analyzed similarly.

\begin{table*}
\caption{Breakdown of library workspace algorithms into the \newalg template for SpGEMM $A_{ij} = \sum_k B_{ik} C_{kj}$.}
\scriptsize
\centering
\begin{tabular}{lccccc}
\toprule
    Library & Insert & Sort & Merge\\
    \midrule
    \multirow{2}{*}{Gustavson's~\cite{Gustavson1978}} & 1-D coordinate list and flag list along $j$, & \multirow{2}{*}{Bucket sort} & \multirow{2}{*}{Boolean indexing} \\
    & deduplication when the flag is true & & \\
    \hline
    \multirow{2}{*}{Cusparse~\cite{demouth2012cusparse}} & 1-D hash table along $j$, & \multirow{2}{*}{Unsorted} & \multirow{2}{*}{Hash table retrieval} \\
    & deduplication when collision & & \\
    \hline
    \multirow{2}{*}{ESC~\cite{Bell2012ESC}} & \multirow{2}{*}{2-D coordinate list per slice of $i$} & \multirow{2}{*}{Lexicographic sort} &  Slice concatenation \\
    & & & Reduce-by-key\\
    \hline
    \multirow{2}{*}{Bulu\c{c}'s\footnotemark~\cite{Buluc2009heap}} & \multirow{2}{*}{Heap with key $(i,j)$ per $k$} & \multirow{2}{*}{Sorted when insert} & Multiway merging\\
    & & & Reduce-by-key\\
    \bottomrule
\end{tabular}
\label{tab:kernel-ism}
\end{table*}

\footnotetext{We list the core ideas of Algorithm I of Bulu\c{c}'s paper, but the actual algorithm computes the multiway merging on the fly.}

\section{Concrete Sparse Workspace Policies}
\label{sec:policy}

As demonstrated in \Cref{sec:prior-ism}, there are many different concrete options for each stage of the \newalg template. A user can synthesize many different concrete workspace policies (or algorithms) 
by mixing different data structure, sorting algorithm, and optimization implementations. This section introduces some of these implementation decisions we made to demonstrate a few new techniques that are compatible with our template. The concrete sparse workspace implementations introduced in this section also give a flavor of the types of workspace optimizations possible with our approach and allow us to evaluate our abstract template concretely in \Cref{sec:eval}.

\subsection{Data Structures}
\label{sec:datastructures}
Our compiler materializes the all array in a COO data structure. We chose this data structure because COO is convenient for transforming the physical organization of sparse tensor components~\cite{guo2023fastrans,smith2015splatt,mueller2020sparse}, and we leverage work on code generation for format conversions between COO and other formats from the literature~\cite{chou:2020:conversion}.

The accumulation array serves as a buffer between the input components and the all array. Sparse iteration scatters components into the accumulation array, and then those componets are sorted and merged with the all array. Therefore, the accumulation array data structure should support efficient deduplication, sorting, and sequential accesses. 
\paragraph{Accumulation Indexing}
A strategy that minimizes data movement is to store the tensor components in an array, where the components are not moved until the accumulation array is freed. Sorting and merging are executed on the indices of each component in the array rather than on the component structs themselves. In this way, only integer indices are moved, reducing memory footprint to about $\frac{1}{M+1}$ where $M$ denotes the mode of input components. During merging, the sorted accumulation indices are used to access the component with the smallest coordinates in the accumulation array. An array structure, however, is not always the best choice; other valid data structures supported by the \newalg template may not support this indexing optimization.

\paragraph{Reallocation}
We can resize the accumulation array after it is merged in order to avoid frequent merging with the all array. 
We use a three-stage piecewise linear function with heuristic thresholds and slopes to determine our allocation size. This heuristic allocation policy avoids memory overallocation when compared to a naive memory reallocation policy, like memory doubling.

\subsection{Sorting Algorithms}
\label{sec:sortingalg}
When the accumulation array reaches its capacity, the indices are sorted based on the coordinates of the tensor components. This computation can be modeled as the sorting of multiple arrays where each corresponds to one level of the result tensor. Although there are various multi-array sorting algorithms (e.g., quick sort, bucket sort, and counting sort~\cite{blelloch1991sort,langr2016aqsort}), we only implement two sorting algorithms that leverage the unique traits of the \newalg sorting stage. In the \newalg arrays, each component has a unique label, represented by a finite positive integer and a known range. For example, for a matrix with shape $(I, J)$, the component with coordinates $(i,j)$ can be labeled using $(i \times J + j)$, and the coordinates of the first mode are in the range $[0, I)$. 
The two sorting algorithms may be combined with each other because they are each defined to sort a single tensor level. 

\paragraph{Bucket Sort} This algorithm sorts the indices of the accumulation array (accumulation id) by a bucket $B$ parametrized by $L$ and $h$. Each bucket holds a list of accumulation indices. $L$ is the length of the bucket, and $h$ is a function that maps the coordinates within a component to an integer (the bucket id) in range $[0, L)$.  
Example mapping functions for the component with coordinates $(i,j)$ of an $I \times J$ matrix include $h(i, j; I, J) = i$ when $L = I$ and $h(i, j; I, J) = (i\times I + j) \% L$ when $L < I$. The bucket sorting algorithm also includes a boolean flag array that records whether an element has been inserted. If $B$[bucket id] is empty, the algorithm will allocate a list at $B$[bucket id] and insert the accumulation id of the inserted component. If $B$[bucket id] is not empty and there is already an accumulation id in $B$[bucket id] whose component has the same coordinates as the one being inserted, the values of the two components are summed. Otherwise, the accumulation id is appended to $B$[bucket id]. 

\paragraph{Coordinate Sort} This algorithm appends inserted accumulation ids to a list, whose length is the capacity of the accumulation array. Unlike the bucket sort algorithm, this sorting algorithm does not reduce values upon insertion. When called by the \newalg template, the coordinate sort applies a 1-D array sort on this list with the comparison function defined as the lexicographic order of each level of coordinates. Like bucket sort, the coordinate sort returns a sorted list of accumulation ids.

\subsection{Optimizations}
\label{sec:optmizations}
We utilize the staging of the \newalg structure to accelerate computation. The insertion stage only writes to the accumulation array at the current time step, which is independent of the sort and merge stages of the previous time step. Therefore, we utilize such independence by pipelining them using multiple threads. Moreover, the all array at the next time step is independent from that of the previous time step. Therefore, we can double buffer the all arrays to avoid auxiliary array creation during merging. Both optimizations are orthogonal to the data structure and sorting algorithm decisions described previously.

\begin{figure}
    \centering
    \begin{minipage}[t]{0.4\textwidth}
        \centering
        \includegraphics[width=0.99\textwidth]{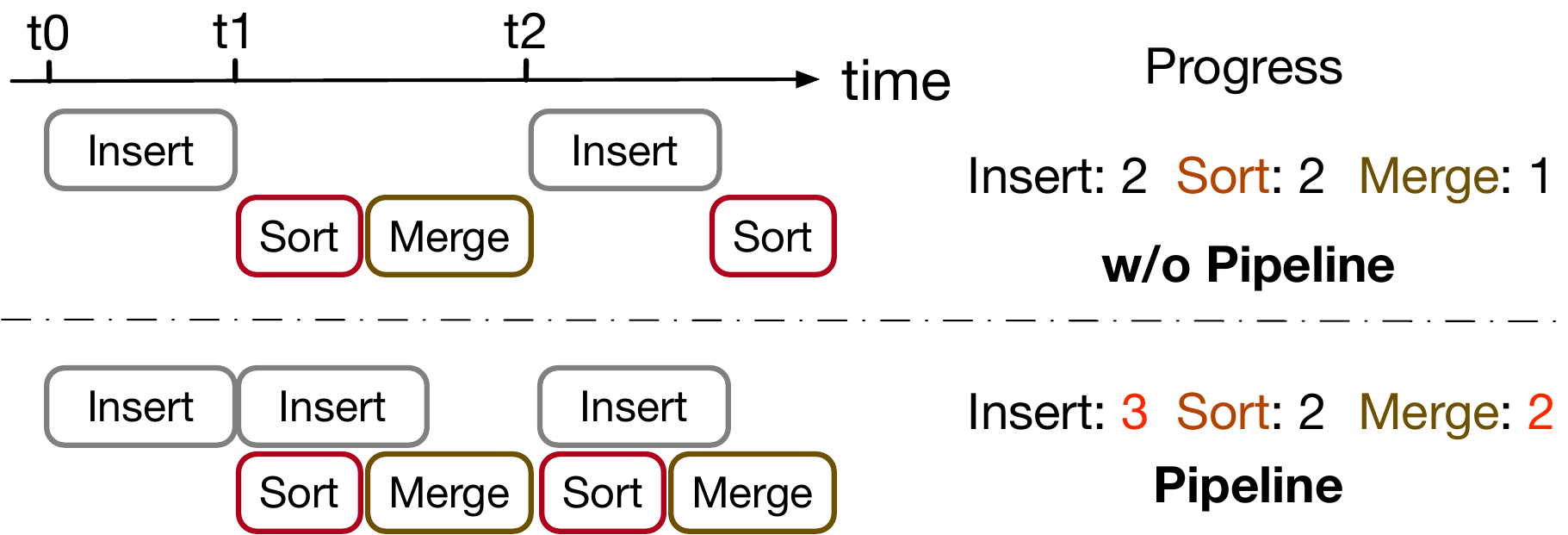}
        \caption{Stage level pipelining of the sort and merge stages with insert.}
        \label{fig:pipeline}
    \end{minipage}%
    \hfill
    \begin{minipage}[t]{0.55\textwidth}
        \centering
        \includegraphics[width=0.99\textwidth]{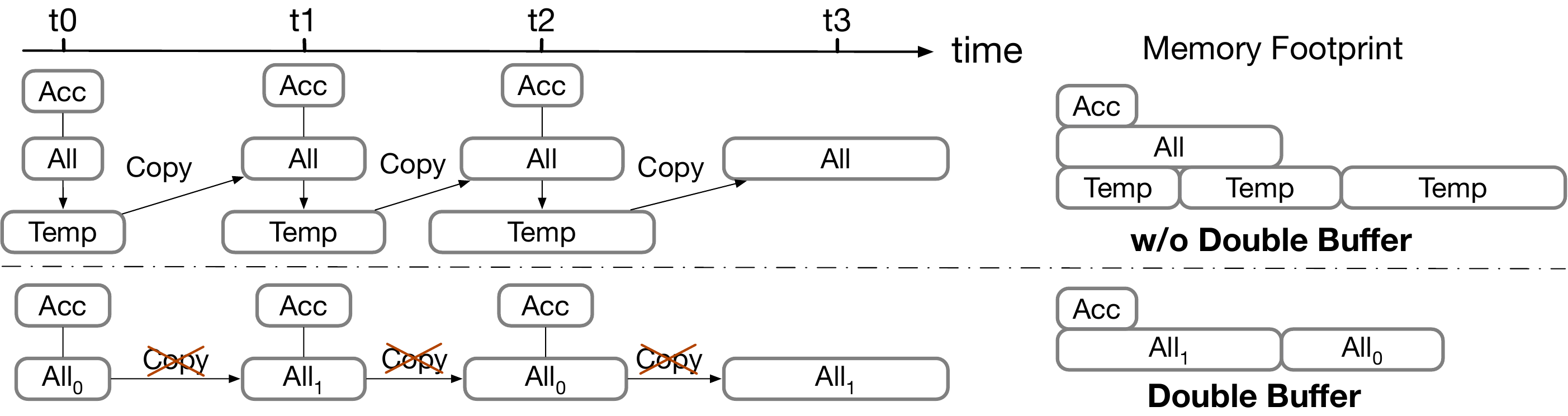}
        \caption{Double buffering the all array, which produces a lower memory footprint.}
        \label{fig:doublebuffer}
    \end{minipage}
\end{figure}

\paragraph{Pipelining (Stage Level Parallelism).} The sort and merge stages can be pipelined with the insertion stage. When the accumulation array is full, our implementation spawns another thread to execute the sort and merge, and the main thread continues to execute the insert for the next sparse iteration step. In this way, the latency of the sort and merge is hidden. However, we now need to allocate two accumulation arrays and switch them between threads when one of them reaches capacity. \Cref{fig:pipeline} shows this---the first accumulation array becomes full at $t_1$, and the second accumulation array is inserted into between $t_2$ and $t_1$ while the first one is being merged concurrently. 

\paragraph{Double Buffering.} Having two copies of the all array can avoid temporary array allocations. We implement the merge step 
as an out-of-place merging of two arrays. If there is only one all array, every merge must allocate a temporary array to hold the merged results before dumping its values into the all array. If we double buffer the all array, then we can eliminate this temporary array. \Cref{fig:doublebuffer} compares these two merging approaches. 

In the ideal case, the execution time of any workspace policy (and the \newalg template) is negligible compared to the total kernel execution time. We provide an ablation study for various policies and compare the empirical results with the lower bound runtime in \Cref{sec:eval-linear}.

\subsection{Towards Parallel Sparse Workspaces}
Although our evaluation is limited to sequential code with sparse workspaces, the \newalglong template is not tied to sequential execution. In general, \newalg enables a decoupled access-compute parallel pattern~\cite{smith1982decoupled} where parallel producers perform sparse iteration and insert components into the accumulation array, while parallel consumers merge the accumulation array with the all array. With proper synchronization strategies between parallel producers and consumers, our framework can be extended to support parallel sparse workspace. The performance challenges lie in how to  increase data locality~\cite{gremse2015gpu}, reduce atomic operations~\cite{nagasaka2017high}, and balance workloads~\cite{anh2016balanced}. Prior kernel-specific work has proposed solutions for these challenges for hand-written kernels~\cite{liu2014bhSPARSE,deveci2017performance,patwary2015parallel,valentin2023spgemm}. The accumulation array can be tiled to better utilize specific architectures~\cite{niu2022tilespgemm}. The all array can be avoided by precomputing the output matrix structure in an extra symbolic phase~\cite{parger2020spECK}. \newalg provides an abstract analysis of these kernels and serves as a foundation for developing code generators that can generate parallel sparse workspaces. We leave exploring the tradeoffs of parallelism-enabled sparse workspace code as future work.

\section{Compilation}
\label{sec:compiler}
This section describes the new compiler techniques of our system.
We first introduce new scheduling and format language commands such that users can manually express tensor expressions with sparse workspaces in our compiler (\Cref{sec:compiler-sched}). To increase productivity, our compiler also has the ability to automatically detect sparse scattering behavior within expressions and then automatically transform those expressions to include missing sparse workspaces for correctness (\Cref{sec:compiler-auto-insert}). Once our compiler detects that a CIN expression contains a sparse workspace tensor, it automatically generates code that inserts the \newalglong algorithm template as described in \Cref{sec:alg-template}.

\subsection{Sparse Workspace Scheduling and Format Commands}
\label{sec:compiler-sched}
Our compiler extends the TACO scheduling language to allow the insertion of sparse workspaces into index notation expressions. We also extend the where statement in CIN, which precomputes an expression into a temporary tensor variable~\cite{kjolstad:2019:workspaces}, to describe parameters that configure the \newalg algorithm for code generation and attributes of the sparse workspace arrays. Users can either directly use the scheduling language to insert sparse workspaces as shown in \Cref{fig:schedule-example-spws}, or rely on our automatic sparse workspace insertion algorithm as shown in \Cref{fig:schedule-example-spws-auto}. 

Users invoke the sparse workspace transformation via the \texttt{precompute} scheduling command~\cite{kjolstad:2019:workspaces}, whose C++ declaration is as follows:
\begin{verbatim}
void IndexStmt::precompute(IndexExpr expr, vector<IndexVar> i_vars, 
                           vector<IndexVar> o_vars, TensorVar ws);
\end{verbatim}
This command transforms a CIN statement that contains a sub-expression \texttt{expr} to a statement with a where sub-statement
$\ldots = \ldots \texttt{ws}_\texttt{o\_vars}~\textbf{where}~ \texttt{ws}_\texttt{i\_vars} = \texttt{expr}$. We call the left-hand side of the where statement the consumer as it consumes the workspace and the right-hand side of the where statement the producer as it produces the workspace data. The producer is a transformed CIN statement with the result tensor of the sub-expression \texttt{expr} replaced by the \texttt{ws}. The consumer assigns the result tensor in the original CIN statement to a transformed expression that uses the \texttt{ws}. The command also optionally replaces the index variables from the original statement \texttt{i\_vars} on the consumer side with the corresponding \texttt{o\_vars}. For simplicity, we assume \texttt{o\_vars} is equivalent to \texttt{i\_vars} for the rest of~\Cref{sec:compiler}. Our compiler's \texttt{precompute} command transformation differs in the tensor variable (\texttt{TensorVar}) type for the workspace and in the construction of the consumer. 

Our additions to the tensor variable format include a sparse format \texttt{SpFormat} class, the output order of the workspace \texttt{ow\_order}, and additional metadata to the \texttt{TensorVar} class. The SpFormat class configures the materialization of the accumulation array. It annotates the specific sorting algorithm by an enum in the SpFormat class (signified by the \texttt{Coord} argument in \Cref{fig:schedule-example-spws}), which also assigns the materialized data structure for the accumulation array. The SpFormat class also annotates the number of orders of the accumulation array, which equals the length of the \texttt{ow\_order}. The \texttt{ow\_order} assigns how the workspace's access order is converted from the producer to the consumer. A concrete algorithm that automatically decides the appropriate \texttt{ow\_order} is described in \Cref{sec:compiler-auto-insert}. The TensorVar metadata stores the dimensions of each level in the all array and the parameters required for the chosen sorting algorithm. For example, the coordinate sort (\texttt{Coord}) algorithm requires the initial capacity of the accumulation array.

These scheduling and format commands expand TACO's scheduling space by introducing additional scheduling options to configure workspaces. They thus alter the performance model of different schedules. Sparse workspaces remove key constraints on scheduled expressions by legalizing sparse scattering behavior. For the performance objectives, sparse workspaces add an intermediate stage to the computation flow, which may favor different types of data locality caused by the user-provided schedules.

\subsection{Automatic Sparse Workspace Insertion}
\label{sec:compiler-auto-insert}
The automatic sparse workspace insertion transformation decides whether or not a sparse workspace should be inserted and, if so, which sparse workspace configurations to use. Users can apply schedules like \texttt{split}, \texttt{pos}, and \texttt{reorder} to a CIN expression as if no sparse scattering behavior exists in the expression. Then, the compiler automatically detects whether sparse scattering occurs and, if necessary, inserts a workspace into the CIN. 

The algorithm deduces the loop order \texttt{input\_order} from the CIN expression and compares it with the access order of the result tensor variable \texttt{output\_order}. If these two orders are the same, then the expression is \textit{concordant}~\cite{ahrens:2022:autoscheduling}. Otherwise, it is \textit{discordant}. If every level of the result tensor supports random insertion and lookups (i.e., it behaves as a dense level format), then we do not insert any workspaces. Even if the expression is discordant, the order mismatch may not require a workspace because it only involves linear transformations on the access order, like $i \times J + j \rightarrow j \times I + i$, which is expressible using dense formats. Otherwise, if the expression is discordant, we assign the dimension of each level of the SpFormat to be the same as the output, and \texttt{ow\_order} is calculated to satisfy the following constraint \texttt{input\_order[i] == output\_order[ow\_order[i]]}. If the expression is concordant, we check the mode formats of all tensors. If the result format has different storage levels than the iteration levels, we insert a sparse workspace and transform the consumer-side assignment to the format conversion IR of~\citet{chou:2020:conversion}. If the result tensor format's storage levels are the same as its iteration levels, we can either insert a workspace with the same levels and dimensions as the result or use the common iteration hoisting optimization (see \Cref{sec:hoist}) to allocate a lower-order sparse workspace.

\paragraph{Input Order Reconstruction.} First, the transformation must deduce the loop order using the original (unscheduled) index variables. The key step in this deduction is to reconstruct the original loop order from the expression by remapping transformed index variables back to their original indices. This retrieval may be complex since the expression may already be composed of other schedules that transform and change the loop index variables without modifying the access index variables~\cite{senanayake:2020:scheduling}. The transformation collects the input-loop index variables \texttt{input\_order} from the composition of $\forall$s in the input expression. Then, our compiler has three rules to retrieve the access index variables from the loop index variables for \texttt{fuse}, \texttt{pos}, and \texttt{split}, respectively. These rules are applied in the inverse order of the scheduling relations in the expression. 
Our compiler targets the following three scheduling commands~\cite{senanayake:2020:scheduling}:

\begin{itemize}
    \item \textbf{split}. \texttt{split(IndexVar i, IndexVar i0, IndexVar i1, size\_t s)} splits an index variable i into two nested index variables (\texttt{i0} and \texttt{i1}) with the iteration step \texttt{s}.
    \item \textbf{fuse}. \texttt{fuse(IndexVar i, IndexVar j, IndexVar f)} fuses two nested index variables (\texttt{i} and \texttt{j}) to an index variable \texttt{f}.
    \item \textbf{pos}. \texttt{pos(IndexVar i, IndexVar p, Access A)} replaces the coordinate space index \texttt{i} with a transformed index variable \texttt{p} that iterates through the position space of input access \texttt{A} over the same iteration range.
\end{itemize}

If a \texttt{split} relation occurs, the transformation detects the locations of the split index variables, \texttt{i0} and \texttt{i1}, in the loop order and determines which is the inner index and which is the outer index position. 
Then, if \texttt{i} is a reduction variable, the transformation replaces the outer split index variable with \texttt{i} and deletes the inner split index variable. Otherwise, it replaces the inner one with \texttt{i} and deletes the outer. This distinction is designed for optimizing common non-reduction iterations. The split transformation does not change the start position of the reduction so the outermost index (between \texttt{i0} and \texttt{i1}) decides the insertion position for \texttt{i}.
If a \texttt{pos} relation occurs, the transformation replaces the index variable \texttt{p} with \texttt{i} since \texttt{pos} does not change the access order.
If a \texttt{fuse} relation occurs, the transformation replaces the index variable \texttt{f} with \texttt{i} and \texttt{j} since \texttt{i} and \texttt{j} were consecutive in loop order when they were fused.

For example, consider an SpGEMM followed by a matrix transposition $A_{ji} = \sum_k B_{ik} C_{kj}$. We assume the user imposes the following schedules on the expression, as shown in \Cref{fig:schedule-example}. The initial value of \texttt{input\_order} equals $\{f0,f1,j\}$. Based on the above rules, we construct the \texttt{input\_order} as $\{f_0,f_1,j\} \Rightarrow \{f_{pos},j\} \Rightarrow \{f,j\} \Rightarrow \{i,k,j\}$.

\begin{figure}
\centering
\scriptsize
\begin{minipage}{0.43\textwidth}
\begin{lstlisting}[language=lang,mathescape=true,basicstyle=\scriptsize\ttfamily,numbers=left,frame=single]
A: ({Compressed,Compressed}, {j,i})$\rightarrow${j,i};
B: ({Dense,Compressed}, {i,k})$\rightarrow${i,k};
C: ({Dense,Compressed},{k,j})$\rightarrow${k,j};
stmt: A(j,i) = B(i,k) * C(k,j);
stmt = stmt.reorder({i,k,j})
            .fuse(i,k,f)
            .pos(f,fpos,B(i,k))
            .split(fpos,{f0,f1},4)
            .reorder({f0,f1,j});
\end{lstlisting}
\vspace{-0.1cm}
\caption{Example input schedule to the Automatic Sparse Workspace Insertion: $\forall_{f_0 f_1 j} A_{ji} \pluseq B_{ik} C_{kj}$ s.t. $fuse(i,k,f)$ and $pos(f,f_{pos},B_{ik})$ and $split(f_{pos},\{f_{0},f_{1}\},4)$}
\label{fig:schedule-example}
\end{minipage} \hfill
\begin{minipage}{0.51\textwidth}
\begin{minipage}{\textwidth}
\begin{lstlisting}[language=lang,mathescape=true,basicstyle=\scriptsize\ttfamily,numbers=left,firstnumber=10,frame=single]
W: Tensor(SpFormat(2, Coord), {I,J}, {1,0}, cap);
stmt = stmt.precompute(B(i,k)*C(k,j), {i,j}, {i,j}, W);
\end{lstlisting}
\vspace{-0.1cm}
\caption{The code required to manually insert a workspace for the example in \Cref{fig:schedule-example}. The updated CIN now contains a \texttt{where} node: $\forall_{j i} A_{ji} =  W_{j i}$ where $\forall_{f_0 f_1 j} W_{i j} \pluseq B_{i k} B_{k j} \ldots$ }
\label{fig:schedule-example-spws}
\end{minipage} 

\vspace{1em}

\begin{minipage}{\textwidth}
\begin{lstlisting}[language=lang,mathescape=true,basicstyle=\scriptsize\ttfamily,numbers=left,frame=single,firstnumber=10]
stmt = insertSparseWorkspace(stmt, Coord, cap);
\end{lstlisting}
\vspace{-1em}
\caption{The automatic sparse workspace insertion command is equivalent to \Cref{fig:schedule-example-spws}.}
\label{fig:schedule-example-spws-auto}
\end{minipage}
\end{minipage}
\end{figure}

\paragraph{Input-output Comparison.} 
This step of the transformation compares the input loop order with the access index variables of the result tensor. First, the transformation identifies the access order of the result tensor \texttt{output\_order}. Then, the transformation eliminates any variables in the \texttt{input\_order} that are not used in the result access. The size of the final \texttt{input\_order} set determines the number of workspace levels and the final \texttt{input\_order} can now be used to compare with \texttt{output\_order} to deduce the correct \texttt{ow\_order}.

For the SpGEMM example, the result tensor access $A_{ji}$ defines \texttt{output\_order} as $\{j,i\}$. The order does not contain a $k$ index so $k$ is a reduction variable. This step eliminates $k$ from the $\texttt{input\_order}=\{i,k,j\}$ to produce the final $\texttt{input\_order}=\{i,j\}.$ The sparse workspace must have two levels with dimensions $\{I,J\}$, which are assembled in the \texttt{input\_order} $\{i,j\}$ and accessed in the \texttt{output\_order} $\{j,i\}$. The \texttt{ow\_order} must be $\left[1,0\right]$ to satisfy the constraints \texttt{input\_order[0] == output\_order[1]} and \texttt{input\_order[1] == output\_order[0]}. The final generated workspace $W$ for this example is shown in \Cref{fig:schedule-example-spws} on line 10.

\paragraph{Common Iteration Hoisting.} 
\label{sec:hoist}
If part of the \texttt{input\_order} and \texttt{output\_order} match, common iteration variables can be hoisted out of the where statement to become nested foralls that serve as shared outer loops for both the consumer and producer. The conditions for this optimization are:
\begin{enumerate}
    \item The output tensor format has the same storage levels as the iteration levels.
    \item The index variables are the same in both orders until one position $p > 0$ where the index variables mismatch. All index variables after position $p$ are ignored.
    \item For all the index variables that match before position $p$, the corresponding modes in the output tensor have stronger abilities than the input tensors.
\end{enumerate}
We measure a tensor's ability using the same method as~\citet{ahrens:2022:autoscheduling}. A tensor’s ability is the combination of the abilities of its level formats. Intuitively, a format's ability measures the time complexity of inserting and accessing the format. If the result format can be assembled using the access pattern generated by the input tensors, we define the result format as stronger than the input. In other words, if the result format is stronger than the input, the result can be directly assembled without the help of a workspace. For example, a dense output is stronger than a compressed input because the dense format supports inserts whereas the compressed format needs to be iterated by position and writes to the result via insertion.

\paragraph{Dense Workspace Optimization.} This optimization is an extreme case of common iteration hoisting, but we define it as a separate case since it allows our compiler to support dense workspaces. If an expression is concordant, it may be identified as having dense scattering behavior, which can be solved by a dense workspace. Dense workspaces may perform better than sparse workspaces in this scenario because the sparse workspace has to store extra index arrays that are not needed by the dense workspace. The conditions for this optimization are:
\begin{enumerate}
    \item There is a reduction variable in the \texttt{input\_order} retrieved from the schedules. In our SpGEMM example, $\{i,k,j\}$ contains the reduction variable $k$.
    \item There is only one variable in the \texttt{input\_order} after the reduction variable. Our outer-product SpGEMM example does not satisfy this constraint. 
\end{enumerate}
If these requirements are met, our transformation uses the dense workspace transformation of~\citet{kjolstad:2019:workspaces}. We analyze the performance of the dense workspace optimization in \Cref{sec:eval-dnws-eigen}.

\subsection{Code Generation}
\label{sec:compiler-code-gen}
After the sparse workspace configurations are set by the scheduling language, our compiler lowers CIN to a C-like imperative IR before finally generating C++ code. The C-like imperative IR models platform-agnostic basic blocks such as variable declarations, loops, conditional statements, and function calls. Our imperative IR introduces a new mechanism for the accumulation and all arrays and generates \newalg as function calls to external functions. During code generation, our compiler integrates \newalg function calls and variables into the generated code, which materializes the specific \newalg policy assigned by the user in the scheduling language.

Our code generation algorithm for sparse workspaces occurs whenever the compiler detects a where statement with a SpFormat tensor variable. The sparse workspace tensor variable is accessed in two different ways on the producer and consumer sides, while an ordinary tensor is only accessible in one way. Our compiler uses the same sparse iteration generation mechanism as TACO, transforming tensor access index variables into per-level iterators~\cite{kjolstad:2017:taco}. The level format of the input tensors determines the attributes and capabilities of the loop iterator. The workspace has the same access attributes as a dense level: random insertion and lookups. Therefore, when generating producer-side co-iteration code, our code generator treats the sparse workspace as a dense tensor and replaces the assignment statement with the \newalg functions.  
The rest of the code generation algorithm follows from the TACO body of work~\cite{chou:2020:conversion,senanayake:2020:scheduling}.
Our compiler modifies code generation in the following steps:

\begin{enumerate}
    \item Our compiler inserts the proper data structure allocations when an SpFormat is detected in the CIN expression. See lines 4--6 in \Cref{fig:spgemm-code} for an example of the generated code. 
    \item To force concordance, our compiler tracks the tensor component's coordinates and reorders them based on the \texttt{ow\_order}. This behavior generates a variable assignment expression in the body of any forall loop that has an index variable participating in the \texttt{output\_order} of the result tensor. See lines 12 and 15 in \Cref{fig:spgemm-code}.
    \item When our compiler detects a where statement with an associated SpFormat workspace tensor, the mechanism forces any loop index variable iterators that are also in the sparse workspace producer access to have dense capabilities. 
    \item Then, our compiler detects the sparse workspace assignment on the producer-side of the where statement and inserts the \newalg function calls. See lines 17--23 in \Cref{fig:spgemm-code}.
    \item Our compiler emits the cleanup \newalg function calls between the producer and consumer sides of the detected where statement. See lines 28--32 in \Cref{fig:spgemm-code}. 
    \item Finally, our compiler lowers the consumer side, which emits the compression function. When generating the compression function, the relevant consumer-side access iterators are created using (higher-order) COO format capabilities. See lines 34--53 in \Cref{fig:spgemm-code}.
\end{enumerate}

The code generation algorithm allows our compiler to fill in the code holes produced by the \newalg template using user-defined functions. We purposefully separate the code generation of the \newalg function template from the IR for modularity. Our template design allows for different optimizations and policies, and we anticipate that users will want to introduce other hand-optimized implementations in the future. Our system is robust to such changes and sufficiently modular to isolate the optimizations to the \newalg function definitions. However, the \newalg template is encoded directly into the IR transformations of our system since we have shown the generality of the \newalg template structure in expressing various policies.

\section{Evaluation}
\label{sec:eval}
We evaluate our sparse workspace technique by comparing the performance of linear and tensor algebra kernels against state-of-the-art systems. We also perform ablation studies on different sparse workspace policies to describe the optimization space of our design.

\subsection{Methodology}
\label{sec:eval-methodology}
All experiments are run on a dual-socket 2.4 GHz Intel Xeon E5-2640v4 machine with 40 cores (80 threads) and 50 MB of L3 cache per socket, running Ubuntu 20.04 LTS. The machine has 256 GB of memory and runs kernel version 5.4.0-155-generic and GCC 9.4.0. 
We compare our work against TACO at commit 2b8ece4~\cite{kjolstad:2019:workspaces}, Eigen version 3.4.0~\cite{guennebaud2010eigen}, and Cyclops Tensor Framework (CTF) at commit e52330f~\cite{singh2022sparsectf}. 
We implement our compiler in C++ as an extension to the TACO compiler.
All experiments are run with 5 warmup rounds and report the arithmetic mean execution time of 20 benchmark rounds.
As in prior work~\cite{SAM}, all SpGEMM kernels are run with the compressed (sparse) matrix operand $B$ and the second operand is that same matrix transposed $B^T$ with the columns shifted by one. We report all runtimes and memory usage in log-scale.

\Cref{fig:eval-taco-dsws-eigen-gust}, \Cref{fig:eval-upperbound}, and \Cref{fig:taco-spws-memory} were run on real-world data from the SuiteSparse matrix collection~\cite{kolodziej2019suitesparse}. To select a representative and diverse set of matrices, we randomly sampled 20 matrices from each category based on the number of rows: less than 12000, 12000 to 180000, and larger than 180000. Tensors in this section that only have their density or mode defined are sparse based on a uniform random distribution.

In this section, a sparse workspace with a ``Bucket'' policy denotes a bucket sort with $L=I$ and $h(i,j)=i$, a ``Hash'' policy denotes a bucket sort with $L<I$ and $h(i,j) = (i * J + j)\% L$, and a ``Coord'' policy denotes the coordinate sort. For the ``Hash'' and ``Coord'', we heuristically assign $L$ as $2^{\left\lceil log_{2}nnz \right\rceil}$, where $nnz$ is the number of nonzeros of the input matrix.
\subsection{First-order Sparse Workspaces in Linear Algebra}
\label{sec:eval-dnws-eigen}
\begin{figure}
\centerline{\includegraphics[width=0.99\textwidth]{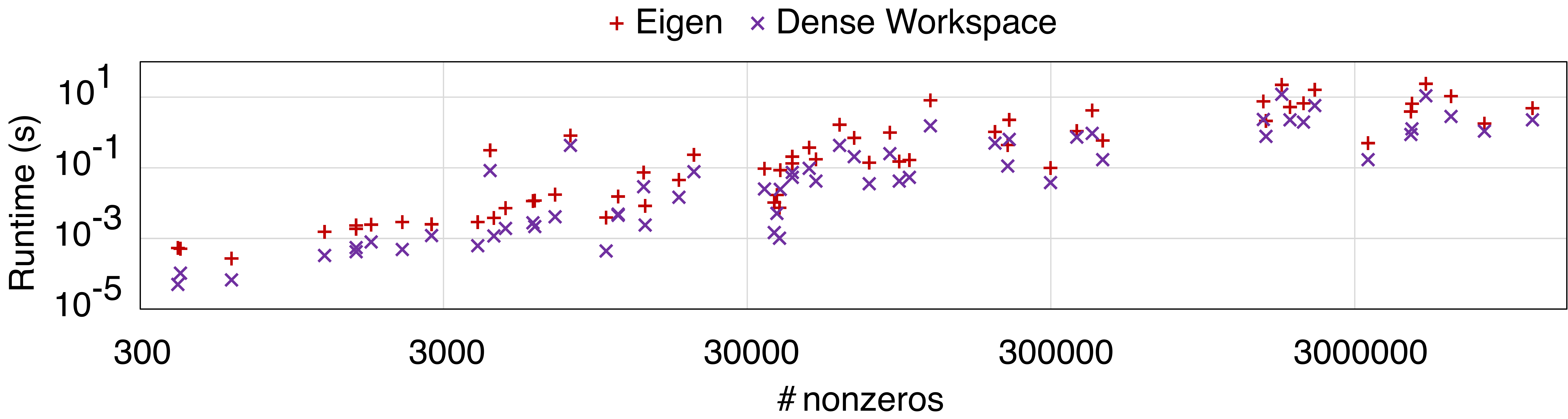}}
\caption{TACO with dense workspace vs. Eigen on row-wise SpGEMM for selected SuiteSparse matrices.}
\label{fig:eval-taco-dsws-eigen-gust}
\end{figure}

Our compiler can generate dense workspace code competitive with a hand-optimized library for first-order (vector) workspaces as expected. In this case, our sparse workspaces are efficiently implemented using the dense workspace technique from prior compilation systems. 
Specifically, in \Cref{fig:eval-taco-dsws-eigen-gust}, we compare against Eigen, a state-of-the-art linear algebra library, for the row-wise SpGEMM with all matrices stored in CSR format. 

\begin{figure}
    \centering
    \includegraphics[width=0.99\textwidth]{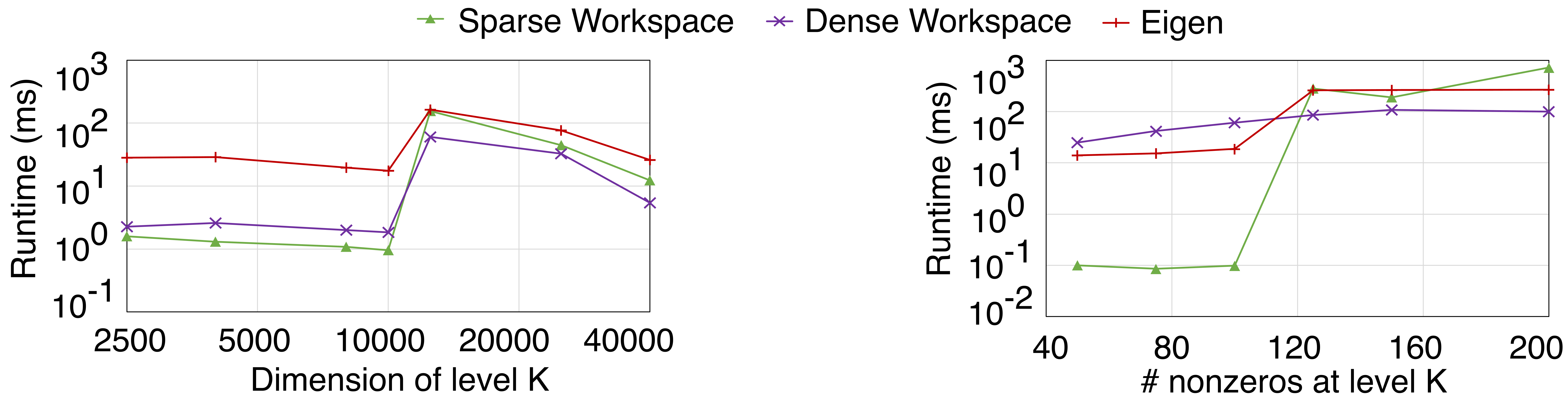}
    \begin{minipage}{0.48\textwidth}
        \centering
        \caption{Comparison of sparse and dense workspaces on row-wise SpGEMM for synthesized matrices.}
        \label{fig:eval-rec-taco-dsws-eigen-gust}
    \end{minipage}%
    \hfill
    \begin{minipage}{0.48\textwidth}
        \centering
        \caption{Comparison of sparse and dense workspaces on outer-product SpGEMM for synthesized matrices.}
        \label{fig:eval-taco-dsws-eigen-outer}
    \end{minipage}
\end{figure}

Though the generated dense workspace kernel outperforms the Eigen library, there exist cases where first-order sparse workspaces, such as those generated by our compiler, outperform both Eigen and dense workspaces. We evaluate the sparse workspace using the same SpGEMM expression with CSR matrices but on synthetic data. In these experiments, we pick the Bucket sparse workspace policy for all linear algebra expressions. The input matrix $B$ is generated with $10 \%$ nonzero elements along level K, and the dimension of level K is swept from 2500 to 40000 with increments generated on a logarithmic base 2 scale. Each column has 1000 nonzeros, and we keep the number of dense elements as $10000\times 10000$. Our synthetic data lets us indirectly control the number of collisions that occur during accumulation, as the dimension of level K increases the amount of coordinate deduplication also increases. As shown in \Cref{fig:eval-rec-taco-dsws-eigen-gust}, the sparse workspace is better than the dense when there is less deduplication. Moreover, first-order dense workspace memory grows proportional to the dimensions of one level. When the amount of deduplication is small, the sparse workspace still consumes less memory because the output tensor components are sparse. However, as the deduplication increases, the dense workspace improves due to better locality.

\subsection{Second-order Sparse Workspaces in Linear Algebra}
\label{sec:eval-linear}

Sparse workspaces use less memory than dense workspaces, which is important when a temporary tensor is very sparse, as a dense workspace may not fit in memory. Moreover, sparse workspaces may improve performance over dense workspaces, due to increased memory locality.

We show the performance and memory benefits of sparse workspaces over the dense baseline on second-order scattering in the outer-product SpGEMM expression. For the compiler-generated algorithms, the input matrices are stored in doubly-compressed sparse row and column (DCSR and DCSC) formats~\cite{Buluc2009heap} since it compresss out the columns without deduplication. 
For Eigen, the input matrices are stored in CSR since it does not support DCSR. The result matrix is stored in CSR so both methods need to compress the workspace to the output format. 

The trend in \Cref{fig:eval-taco-dsws-eigen-outer} is similar to that of \Cref{fig:eval-rec-taco-dsws-eigen-gust}. Again, the sparse workspace is better when there is less deduplication, but the second-order workspace has greater overall speedup since higher orders have a larger opportunity for savings. Unlike \Cref{fig:eval-rec-taco-dsws-eigen-gust}, we keep the dimensions of I and K as 10000 and 10000, respectively, for \Cref{fig:eval-taco-dsws-eigen-outer} and sweep the number of nonzeros at level K. We still keep 1000 elements in each column with nonzeros in matrix $B$.

\begin{figure}
\centerline{\includegraphics[width=0.99\textwidth]{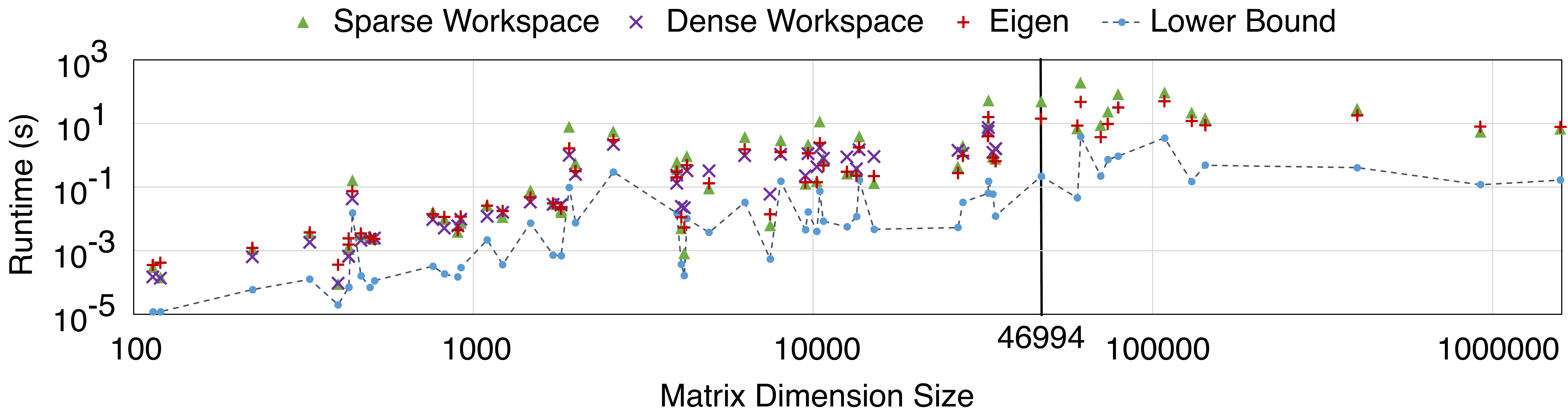}}
\caption{Comparison among the best results of different sorting algorithms with the dense workspace on second-order scattering. We construct the lower bound as described in \Cref{sec:policy}.}
\label{fig:eval-upperbound}
\end{figure}

As shown in \Cref{fig:eval-upperbound}, the performance of sparse workspaces remains competitive even where a dense workspace fits in memory. Also, the sparse workspace only incurs overhead proportional to the number of sparse iterations because the sparse workspace scales at the same rate as the runtime lower bound. Following the definition of the runtime lower-bound in \Cref{sec:optmizations}, we record the ideal execution time by only keeping the sparse iteration computation and eliminating all the code related to the workspace.

\begin{figure}
\centerline{\includegraphics[width=0.99\textwidth]{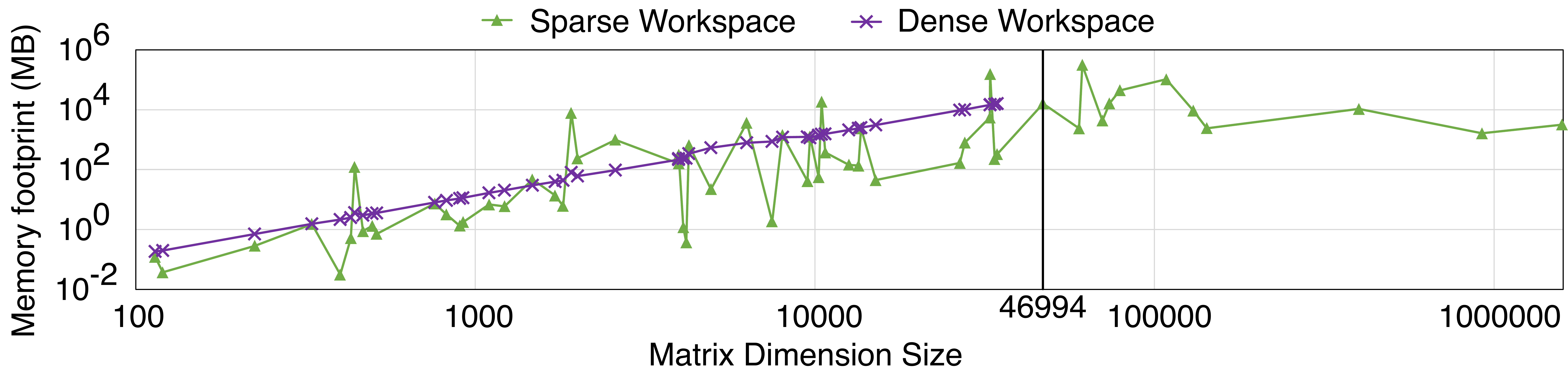}}
\caption{The memory footprint of sparse vs. dense workspaces on second-order scattering.}
\label{fig:taco-spws-memory}
\end{figure}

A sparse workspace is necessary for sparse scattering when a dense workspace is too large to fit in memory. As shown in \Cref{fig:taco-spws-memory}, the sparse workspace also saves on average 3.6$\times$ the amount of memory for the matrices that do not OOM.  We estimate the memory footprint as $\textit{shape} \times (3 \times 4 + 1)$ bytes for the dense workspace and $\mathit{nnz} \times 3 \times 4$ bytes for the sparse workspace, where $\mathit{shape}$ is the number of dense elements, and $\mathit{nnz}$ is the number of nonzeros. We confirm this trend in \Cref{fig:taco-spws-memory}, which demonstrates that the dense workspace runs out of memory when the input matrices get too large (after the black line), while the sparse workspace scales. 

\subsection{Sparse Workspaces in Tensor Algebra}
Sparse workspaces are especially important for higher-order tensor algebra expressions, as they tend to need higher-dimensional temporaries. We demonstrate this effect for SpMTTKRP $\forall_{klij} A_{ij} = B_{kli} C_{kj} D_{lj}$, SpTTM  $\forall_{kijl} A_{ijl} = B_{kij} C_{kl}$ and SpTTM-I $\forall_{kijl} A_{kjl} = B_{kij} C_{il}$, which are used to factorize tensors~\cite{li2016spttm,smith2017spmttkrp}. 
We test these expressions on the nell-2, uber3, and nips3\footnote{We modify the FROSTT uber and nips 4-tensors to 3-tensors by dropping one dimension as in~\citet{hellsten2023baco}.} from the FROSTT dataset~\cite{frosttdataset}, and the Facebook tensor~\cite{2009facebook}  that fit in our memory.

\begin{figure}
    \centering
    \includegraphics[width=0.99\textwidth]{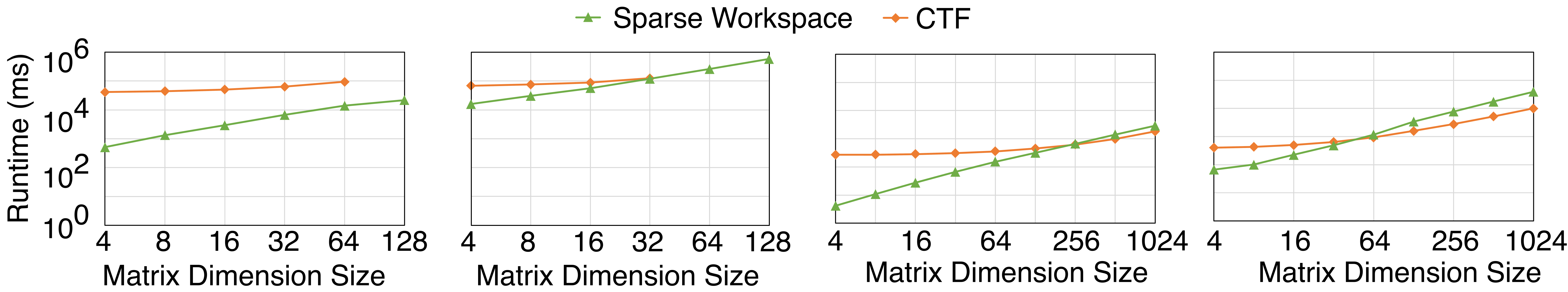}
    \hfill
    \begin{subfigure}{0.24\textwidth}
        \centering
        \caption{SpMTTKRP on nell-2.}
        \label{fig:eval-mttkrp-nell2}
    \end{subfigure} \hfill
    \begin{subfigure}{0.24\textwidth}
        \centering
        \caption{SpTTM on nell-2.}
        \label{fig:eval-ttm-nell2}
    \end{subfigure}
    \hfill
    \begin{subfigure}{0.24\textwidth}
        \centering
        \caption{SpMTTKRP on uber3.}
        \label{fig:eval-mttkrp-uber3}
    \end{subfigure} \hfill
    \begin{subfigure}{0.24\textwidth}
        \centering
        \caption{SpTTM on uber3.}
        \label{fig:eval-ttm-uber3}
    \end{subfigure}
    \caption{SpMTTKRP and SpTTM runtime on nell-2 and uber3.}
    \label{fig:eval-higher-order}
\end{figure}

As shown in \Cref{fig:eval-higher-order}, we compare the generated Hash sparse workspace to CTF, a state-of-the-art sparse tensor algebra library~\cite{singh2022sparsectf}. We sweep the dimensions of the free level of the matrix (\texttt{J} for SpMTTKRP and \texttt{L} for SpTTM) and keep the density of the input matrices as $10\%$. For both systems, we store input higher-order tensors in the compressed sparse fiber (CSF)~\cite{smith2015csf} format and output tensors as COO. The input matrices to CTF are stored in COO since that is the only format it supports for matrices. CTF reduces higher-order tensor contractions to matrix multiplications through index folding, which costs time and memory. Therefore, CTF runs out of memory in \Cref{fig:eval-ttm-nell2} and \Cref{fig:eval-mttkrp-nell2}. Also, CTF computes some metadata of the tensors for optimization before computation. Such overhead is apparent when the matrix is small. When the matrix grows larger, the benefits from the metadata outweigh the latency of the preprocessing. Therefore, in \Cref{fig:eval-mttkrp-uber3} and \Cref{fig:eval-ttm-uber3} the runtime of the sparse workspace grows faster than CTF as the matrix dimension increases. We also evaluate SpMTTKRP on the freebase\_sampled tensor from Freebase~\cite{2015freebase} with $\texttt{J}=4$. CTF OOMs in this case, while our method takes around 16 hours to finish.

\begin{figure}
    \centering
    \includegraphics[width=0.99\textwidth]{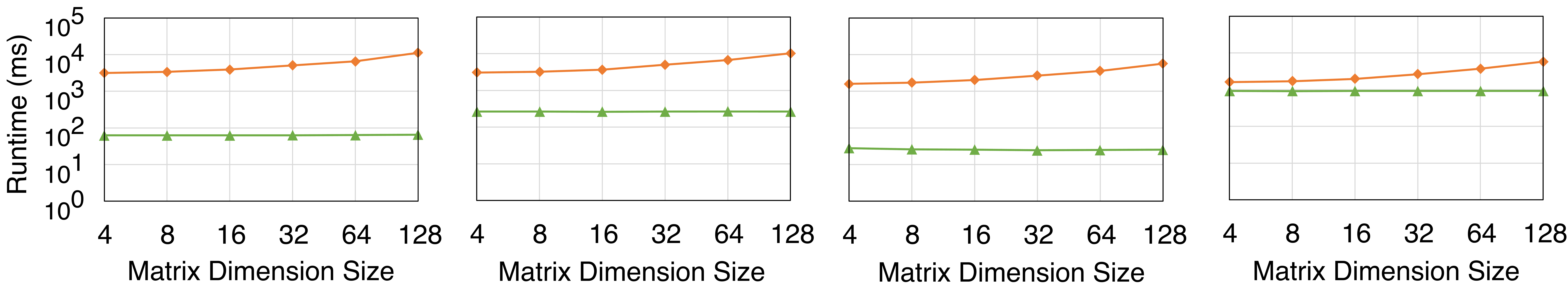}
    \hfill
    \begin{subfigure}{0.24\textwidth}
        \centering
        \caption{SpMTTKRP on fb.}
        \label{fig:eval-mttkrp-facebook}
    \end{subfigure} \hfill
    \begin{subfigure}{0.24\textwidth}
        \centering
        \caption{SpTTM-I on fb.}
        \label{fig:eval-ttm-facebook}
    \end{subfigure}
    \hfill
    \begin{subfigure}{0.24\textwidth}
        \centering
        \caption{SpMTTKRP on nips3.}
        \label{fig:eval-mttkrp-nips3}
    \end{subfigure} \hfill
    \begin{subfigure}{0.24\textwidth}
        \centering
        \caption{SpTTM-I on nips3.}
        \label{fig:eval-ttm-nips3}
    \end{subfigure}
    \caption{SpMTTKRP and SpTTM-I runtime on Facebook (fb) and nips3.}
    \label{fig:eval-higher-order-2}
\end{figure}

To show that the performance of sparse workspaces is agnostic to the shape of the input tensors, we do experiments in \Cref{fig:eval-higher-order-2}. Unlike \Cref{fig:eval-higher-order}, we keep the input non-zero elements unchanged and sweep the dimensions of the free level of the matrix. The other experimental settings are the same as \Cref{fig:eval-higher-order}. As expected, the runtime of the sparse workspace does not increase as the matrix dimension size becomes larger since the sparse iteration and \newalg behaviors do not change. On the contrary, the runtime of CTF increases because it uses dense sub-tensors to support its index folding, whose cost grows with the size of the matrix.

\subsection{Study of Sparse Workspace Design Choices}
\label{sec:eval-design-choices}
We perform an ablation study to analyze the different optimizations and concrete policies of our sparse workspace generation framework. We first empirically justify the two-level accumulation and all arrays, instead of only using one array as in prior work policies~\cite{Amir2022semiring,scott2023index}. We then investigate the influence of the concrete policies introduced in \Cref{sec:policy}.

\begin{figure}
\centerline{\includegraphics[width=0.99\textwidth]{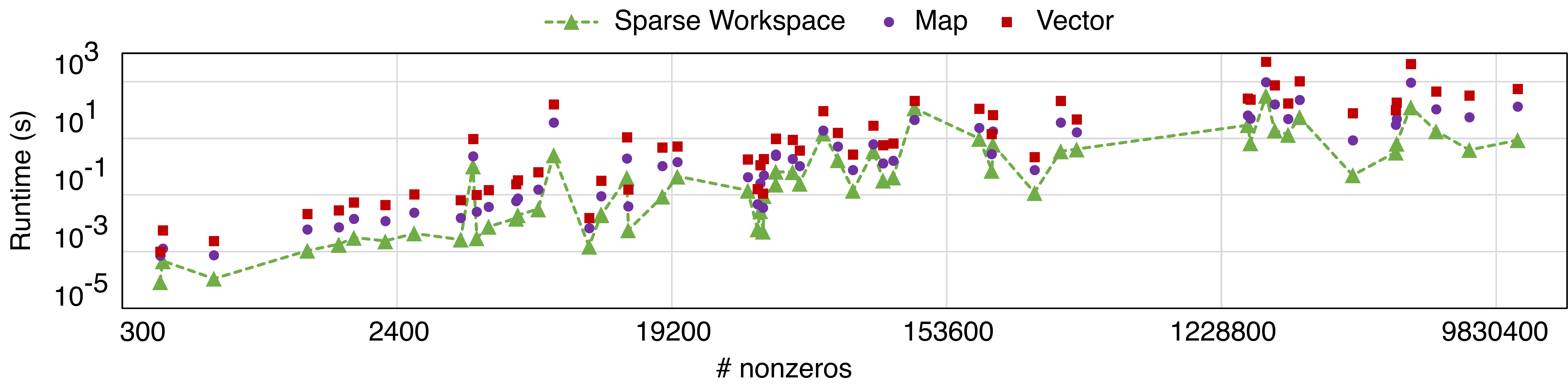}}
\caption{Empirical benefits of the accumulation and all array structure in our \newalg template. We compare our bucket policy against a map and vector data structure policy. The map implicitly sorts every insert (accumulation size = 1), whereas the vector only sorts once before compressing the output (accumulation size = sizeof(output)), which are the two extremes of our \newalg template.  }
\label{fig:eval-noacc}
\end{figure}

In \Cref{sec:two-array}, we analyzed the benefit of using a two-level array structure via Big-O analysis. However, since Big-O can be misleading in practice, we compare our optimized methods with some straightforward sparse workspace policies that only use one storage array. Our two-array workspace policies have two extremes. One extreme sorts and deduplicates every time an output component is inserted, which occurs when using a single map as a sparse workspace~\cite{Amir2022semiring}. The other extreme is to sort and deduplicate after all output components are generated, which occurs when inserting components into a single vector with one sorting and deduplication pass at the end before compression. In both of these extremes, only a single data structure is necessary. Our proposed method lies in the middle because we sort and deduplicate in batches using the accumulation array and merge the components with the all array every time the accumulation array is full. Though the two extremes do not need a merging stage, all three methods require a compression stage. \Cref{fig:eval-noacc} shows that these two extremes---the map and the vector---are slower than our bucket policy in most cases. Therefore, it is empirically beneficial to use our \newalg algorithm template with two arrays.

\begin{figure}
\centerline{\includegraphics[width=0.99\textwidth]{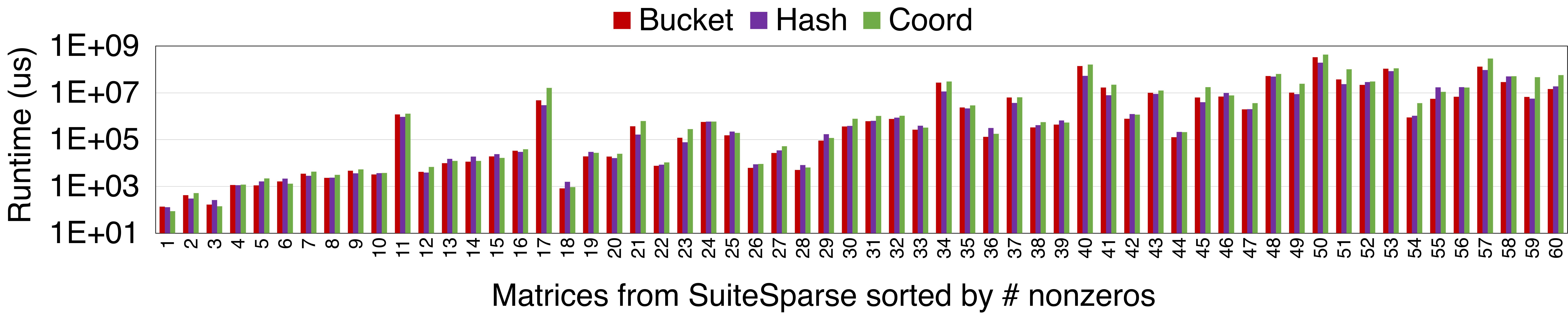}}
\caption{Runtime comparison among different sorting algorithms.}
\label{fig:eval-sorting}
\end{figure}

\begin{figure}
\centerline{\includegraphics[width=0.99\textwidth]{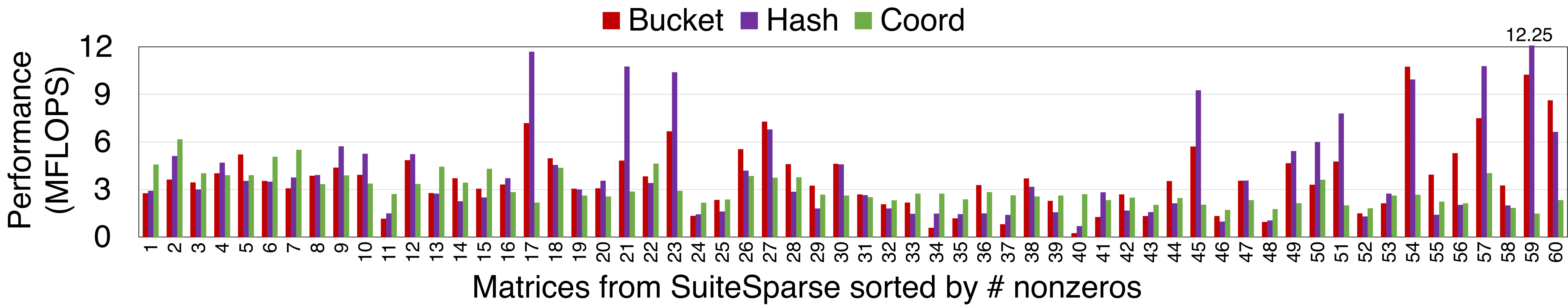}}
\caption{Performance comparison among different sorting algorithms reported in a linear scale. FLOPS is calculated as the number of fused multiply-add (\texttt{fma}) operations divided by the runtime.}
\label{fig:eval-sorting-flops}
\end{figure}

\begin{figure}
    \centering
    \includegraphics[width=0.99\textwidth]{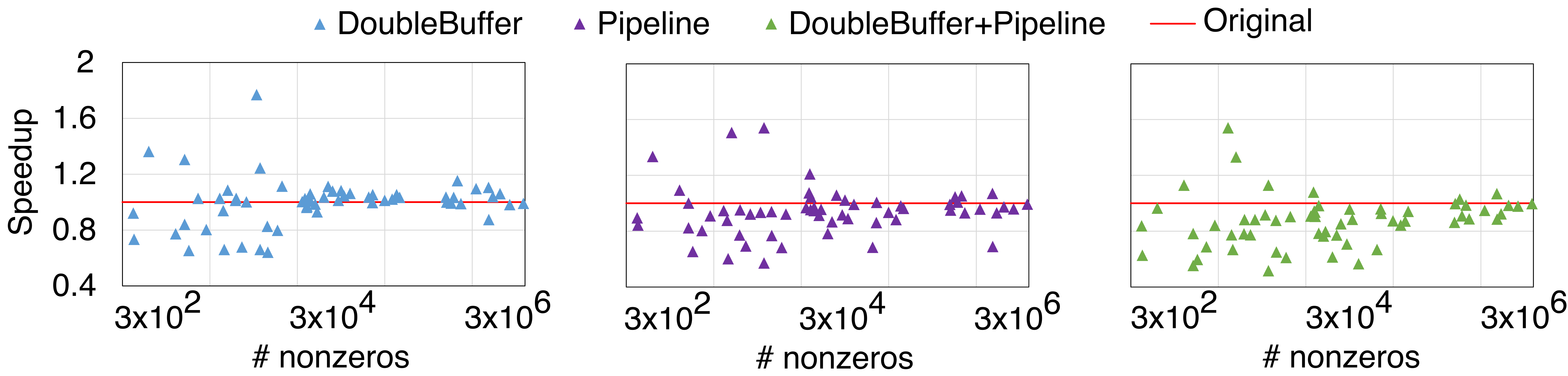}
    \begin{subfigure}{0.33\textwidth}
        \centering
        \caption{Double Buffer.}
        \label{fig:eval-hash-doublebuffer}
    \end{subfigure}%
    \begin{subfigure}{0.33\textwidth}
        \centering
        \caption{Pipeline.}
        \label{fig:eval-hash-pipeline}
    \end{subfigure}%
    \begin{subfigure}{0.33\textwidth}
        \centering
        \caption{Double Buffer + Pipeline.}
        \label{fig:eval-hash-db+pipeline}
    \end{subfigure}    
    \caption{Bucket sort using two optimizations proposed in \Cref{sec:optmizations}. Speedups are reported in a linear scale.}
    \label{fig:eval-optimizations}
\end{figure}

In \Cref{sec:sortingalg} and \Cref{sec:optmizations}, we introduced two concrete sorting algorithms and two optimizations for sparse workspaces that fit within our \newalg framework. \Cref{fig:eval-sorting}, \Cref{fig:eval-sorting-flops}, and \Cref{fig:eval-optimizations} show that no single sorting algorithm or optimization can dominate all inputs. As shown in \Cref{tab:eval-mttkrp-ablation}, this observation still holds for higher-order tensors.  In \Cref{tab:eval-mttkrp-ablation}, we sweep the dimensions of the free level \texttt{J} for SpMTTKRP and keep the density of the input matrices as $1\%$. Other tensors' results can be found in the appendix, and they share similar trends.  

Since the performance of sparse computation is data-dependent, the compiler should support many concrete sparse workspace policies to maintain performance across the range of sparse data.
\begin{table*}
\caption{SpMTTKRP runtime (ms) of sorting algorithms on nips3. We underscore the best policy for each \texttt{J}.}
\scriptsize
\centering
\begin{tabular}{lcccccccccccc}
\toprule
    Method & J=4 & J=8 & J=16 & J=32 & J=64 & J=128 & J=256 & J=512 & J=1024 & J=2048\\
    \midrule
    Bucket & 0.4390 & 1.037 & 3.022 & 7.904 & 27.41 & 72.40 & 165.3 & 322.1 & \underline{664.6} & \underline{1331} \\
    Hash & 0.5954 & 1.204 & 3.139 & \underline{7.576} & \underline{26.08} & \underline{64.65} & \underline{152.3} & \underline{317.9} & 673.5 & 1410 \\
    Coord & \underline{0.3422} & \underline{0.9175} & \underline{2.883} & 7.687 & 27.18 & 71.5 & 163.0 & 336.7 & 706.7 & 1388 \\
    \bottomrule
\end{tabular}

\label{tab:eval-mttkrp-ablation}
\end{table*}

\section{Related Work}
\label{sec:related-work}
We introduce four areas of related work. The first area, code composition in compilers, describes our work in terms of general compilation strategies from prior work. Then, we focus specifically on works related to workspaces and sparse systems, namely workspace optimizations in sparse kernels, workspace optimizations in sparse compilers, and sparse tensor format abstraction.

\paragraph{Code Composition in Compilers}
Code composition enables the modular reuse of high-performance codelets (or stencils) without modifying the compiler cross-layer. Exocompilation~\cite{exocompilation} provides a framework that replaces computation procedures with hardware primitive functions. Mosaic and TVM~\cite{bansal2023mosaic,chen2018tvm} integrate vendor library functions into their DSL lowering process. Template JITs~\cite{vcode,ertl2004jit,xu2021copy,pichler2023hybrid} apply a similar methodology to general compilation. However, they integrate pre-defined bytecodes or AST nodes instead of library functions. Our method is most similar to Hyper~\cite{neumann2011efficiently}, FFTW~\cite{frigo1998fftw}, and the ideas in~\cite{stichnoth1997composition} in that the compiler composes user-defined algorithmic templates with the rest of the generate code.

\paragraph{Workspace Optimizations in Sparse Kernels}
Prior work on individual tensor algebra kernels has used workspaces for both dense and sparse scattering. The dense workspace was first used to implement the multiple-switch algorithm in SpGEMM~\cite{Gustavson1978}. Other work either generalizes the dense workspace to more expressions~\cite{john1992spa,azad2017work} and/or to hardware architectures~\cite{dalton2015CUSP,yusuke2018spgemm}. A sparse workspace was first proposed in~\cite{Buluc2009heap} to scale SpGEMM to thousands of processors. Subsequent work also generalizes sparse workspaces to more expressions~\cite{akbudak2014spref,ordonez2016gamma} and various architectures~\cite{Bell2012ESC,pal2018outerspace,yang2023ISOSceles}.

Current sparse library frameworks also leverage workspaces. Eigen~\cite{guennebaud2010eigen} uses optimized Gustavson's dense workspace by converting the input matrices to be compatible with row-wise SpGEMM. CTF~\cite{singh2022sparsectf} reduces tensor contractions to matrix multiplications and transpositions where each sub-tensor result works as a workspace that is reduced in the end.

Unlike these prior workspace algorithms and library frameworks, our compiler approach is general for any tensor algebra expression and modular across various workspace designs. The design of our compiler allows for optimization techniques and core data structure strategies while abstracting away architecture-specificity.

\paragraph{Workspace Optimizations in Sparse Compilers}
This paper integrates the sparse workspace algorithm into an existing sparse compilation framework, enabling it to generate code for arbitrary tensor algebra expressions that have sparse scattering.
Early work on sparse compilers only had first-order and second-order dense workspaces for linear algebra. Bik and Wijshoff~\cite{bik1994automatic} materialize the internal sparse collection as a dense workspace in their restructuring compiler for row-wise and outer-product SpGEMM. In SIPR~\cite{william1998sipr}, a dense workspace is implemented for row-wise SpGEMM as a C++ class with access and update methods. Recent work on sparse compilers generalize dense workspaces to tensor algebra~\cite{scott2023index,bik:2022:mlirsparse,kjolstad:2019:workspaces} using schedules~\cite{halide,chen2018tvm,senanayake:2020:scheduling,kjolstad:2019:workspaces}.
Workspaces are indispensable for the scattering problem, therefore, sparse compilers without workspaces can only support appending expressions~\cite{zhao2023spf,zheng2022sparta,ye2023sparsetir}. 

\paragraph{Sparse Tensor Format Abstraction}
The sparse workspace techniques in this work bridge the gap between prior techniques on sparse iteration and tensor format conversions to solve the sparse scattering problem. Without format conversions, the sparse workspace would be unable to generate the result tensor format assigned by the user.
Researchers have been studying sparse tensor formats and efficient format conversion algorithms for decades in order to utilize the data distribution for optimizing specific expressions~\cite{Gustavson1978,saad2003dia,wang2016dok,kincaid1989ell,Buluc2009heap,winter2019adaptive}. Early sparse compilers have their own individual format systems~\cite{thibault1994format,kotlyar1997relational,arnold2010ll}, but recent work on compilers generalize these approaches by creating a uniform representation and conversion routine for common compressed formats~\cite{chou2018format,mueller2020sparse,chou:2020:conversion}. The sparse workspace tensor format described in our work (SpFormat) for \newalg arrays is defined using the same level format abstraction presented by~\citet{chou2018format}, and the code generation for the compression stage of our compiler leverages work from~\citet{chou:2020:conversion}.
 
\section{Conclusion}
\label{sec:conclusion}

We introduce a compiler for sparse tensor algebra that can generate code for expressions with sparse scattering behavior through the introduction of sparse workspaces. Users may either manually insert sparse workspaces using our compiler or rely on the compiler to automatically detect sparse scattering and insert these workspaces. The compiler leverages a four-stage algorithm template, called insert-sort-merge, to abstractly represent a sparse workspace, and we provide data structures and optimizations that implement concrete sparse workspace policies using this algorithm template. 
Our work extends the generality of prior compilers for sparse tensor algebra by treating sparse tensors as a first-class concept for any tensor variable, including temporaries.
\begin{acks}
\label{sec:ack}
We thank Alexander J. Root, Benjamin Driscoll, Bobby Yan, Chris Gyurgyik, David Broman, Guohao Dai, Kai Zhong, James Dong, Praneeth Kolichala, Rubens Lacouture, Scott Kovach, Shiv Sundram, Shulin Zeng, Rohan Yadav, Yu Wang and Zhenhua Zhu for
their helpful feedback. This work was supported in part by the Semiconductor Research Corporation (SRC) PRISM center.
Olivia Hsu was supported by an NSF GRFP Fellowship. Any opinions,
findings, and conclusions or recommendations expressed in this material are those of the authors
and do not necessarily reflect the views of the aforementioned funding agencies.
\end{acks}



\bibliographystyle{ACM-Reference-Format}

\bibliography{references}
\clearpage
\appendix
\section*{Appendix}
\section{Big-O Analysis of the Accumulation Array}
\label{sec:bigo-appendix}

Using just a single memory pool (i.e. with just an all array), the cost of insertion and sorting for each tensor component is 
{\footnotesize $$\sum_{i=1}^{N} O(\sum_{j=1}^{i}j log(\sum_{j=1}^{i}j))= \sum_{i=1}^{N} O(i^2 log(i^2))=O(N^3logN)$$} where $N$ is the number of output nonzeros if the all array is sorted every time it is inserted, or 
{\footnotesize $$  O(\sum_{j=1}^N j log \sum_{j=1}^N j)=O(N^2 log N)$$} if the all array is sorted once. However, assuming the accumulation array is $K$ times smaller than the all array, and the capacity does not change, the insertions into the all array are batched by the capacity of the accumulation array $\Tilde{N}=N/K$, changing the cost to construct the output datastructure to 
{\footnotesize $$\sum_{k=1}^{K}\sum_{i=1}^{\Tilde{N}} O(\sum_{j=1}^{i} j log \sum_{j=1}^{i} j)=\sum_{k=1}^{K} \sum_{i=1}^{\Tilde{N}} O(i^2 log(i^2))= O (\Tilde{N}^2 log \Tilde{N}),$$}
which is smaller than single memory pool by a constant factor.

\section{Additional SpMTTKRP Performance Comparison}
\label{sec:spmttkrp-ablation-appendix}
\begin{figure}[h]
\centerline{\includegraphics[width=0.99\textwidth]{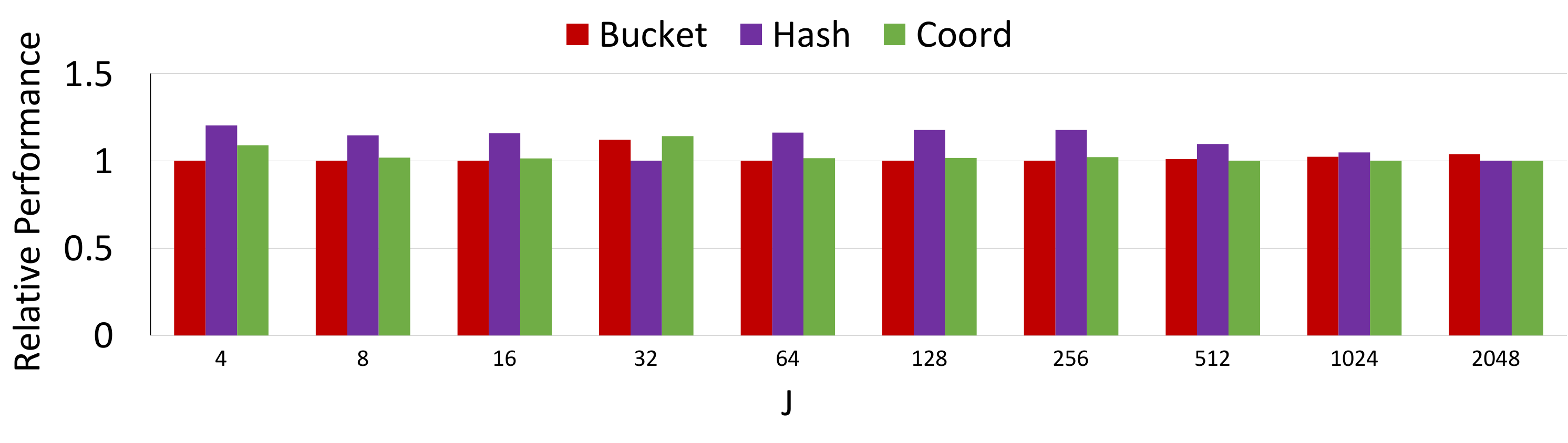}}
\caption{facebook}
\label{fig:facebook-append}
\end{figure}

\begin{figure}[h]
\centerline{\includegraphics[width=0.99\textwidth]{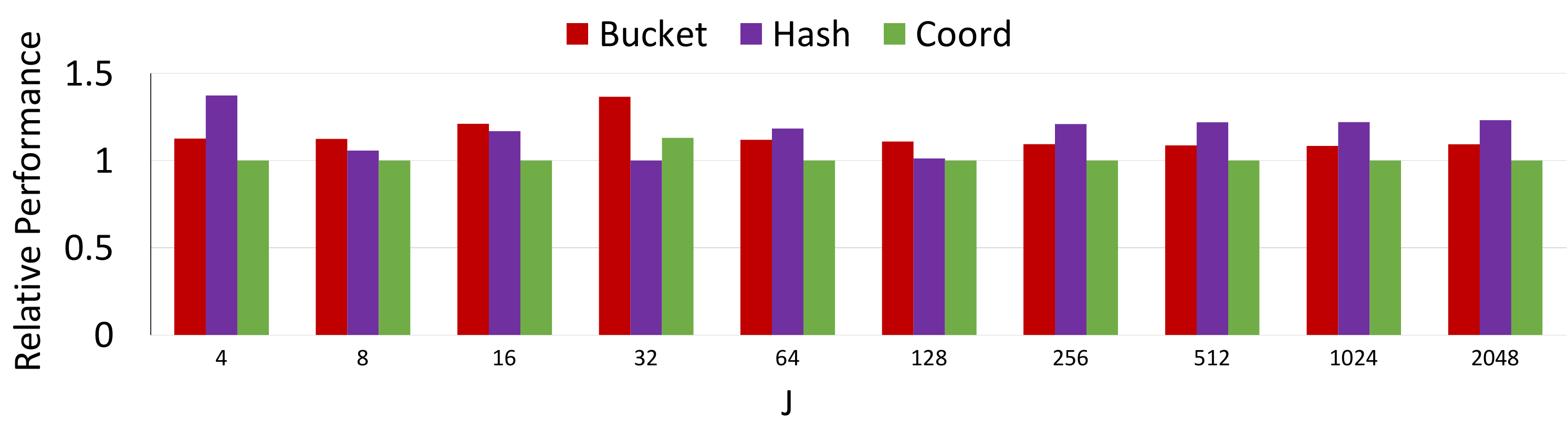}}
\caption{uber3}
\label{fig:uber3-append}
\end{figure}

\begin{figure}[ht]
\centerline{\includegraphics[width=0.99\textwidth]{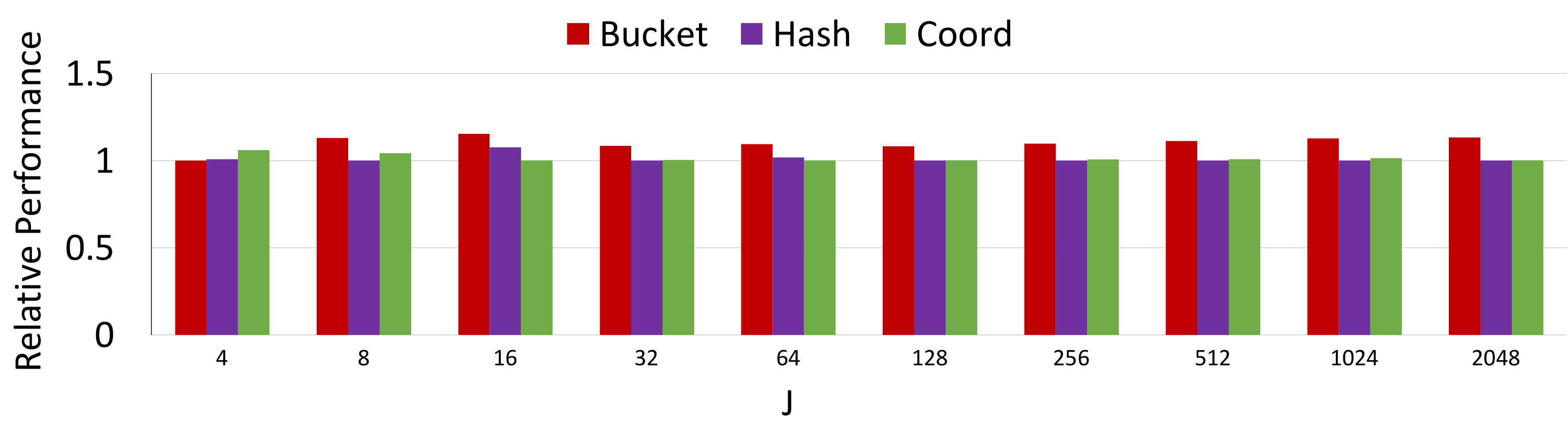}}
\caption{Performance of various sparse workspace policies using nell-2}
\label{fig:nell2-append}
\end{figure}

\Cref{fig:facebook-append}, \Cref{fig:uber3-append}, and \Cref{fig:nell2-append} show SpMTTKRP runtime among different sorting algorithms on facebook, uber3, and nell-2 tensors, respectively.

\section{Tensor Dataset Information}

\begin{table*}[h]
\caption{Information on the selected tensors used to evaluate our work in \Cref{sec:eval}.}
\scriptsize
\centering
\begin{tabular}{lcccl}
\toprule
    Name & Dimension & NNZ & Sparsity & Dataset\\
    \midrule
    uber3 & 183 $\times$ 24 $\times$ 1140 & 3309490 & 0.660988480 & FROSTT~\cite{frosttdataset}\\
    nell-2 & 12092 $\times$ 9184 $\times$ 28818 & 76879419 & 0.000024022 & FROSTT~\cite{frosttdataset}\\
    nips3 & 2482 $\times$ 2862 $\times$ 14036 & 3101609 & 0.000311080 & FROSTT~\cite{frosttdataset}\\
    facebook & 1504 $\times$ 42390 $\times$ 39986 & 737934 & 2.895E-7 & Facebook~\cite{2009facebook}\\
    freebase\_sampled & 38954435 $\times$ 38955429 $\times$ 532 & 139920771 & 1.733E-10& Freebase~\cite{2015freebase}\\
    \bottomrule
\end{tabular}

\label{tab:appendix-tensor-info}
\end{table*}

\begin{table*}
\caption{Information on the selected matrices from SuiteSparse~\cite{kolodziej2019suitesparse} used in \Cref{sec:eval}.}
\scriptsize
\centering
\begin{tabular}{lcccc}
\toprule
    Name & Dimension & NNZ & Sparsity & Kind\\
    \midrule
gams10am & 114 $\times$ 171 & 407 & 0.020878219 & linear programming problem \\
ch5-5-b4 & 120 $\times$ 600 & 600 & 0.008333333 & combinatorial problem \\
lpi\_cplex2 & 224 $\times$ 378 & 1215 & 0.01434949 & linear programming problem \\
lp\_scfxm1 & 330 $\times$ 600 & 2732 & 0.01379798 & linear programming problem \\
odepb400 & 400 $\times$ 400 & 399 & 0.00249375 & 2D/3D problem \\
oscil\_dcop\_39 & 430 $\times$ 430 & 1544 & 0.00835046 & subsequent circuit simulation problem \\
oscil\_dcop\_30 & 430 $\times$ 430 & 1544 & 0.00835046 & subsequent circuit simulation problem \\
ex2 & 441 $\times$ 441 & 13640 & 0.070135386 & computational fluid dynamics problem \\
lp\_standmps & 467 $\times$ 1274 & 3878 & 0.006518107 & linear programming problem \\
west0497 & 497 $\times$ 497 & 1727 & 0.006991648 & chemical process simulation problem \\
gre\_512 & 512 $\times$ 512 & 2192 & 0.008361816 & directed weighted graph \\
fs\_760\_3 & 760 $\times$ 760 & 5976 & 0.01034626 & subsequent 2D/3D problem \\
bp\_1400 & 822 $\times$ 822 & 4790 & 0.007089113 & subsequent optimization problem \\
pde900 & 900 $\times$ 900 & 4380 & 0.005407407 & 2D/3D problem \\
n4c6-b14 & 920 $\times$ 6300 & 13800 & 0.002380952 & combinatorial problem \\
lp\_woodw & 1098 $\times$ 8418 & 37487 & 0.004055734 & linear programming problem \\
fpga\_dcop\_36 & 1220 $\times$ 1220 & 5892 & 0.003958613 & subsequent circuit simulation problem \\
bcsstk12 & 1473 $\times$ 1473 & 17857 & 0.00823006 & duplicate structural problem \\
freeFlyingRobot\_3 & 1718 $\times$ 1718 & 6975 & 0.002363186 & optimal control problem \\
adder\_dcop\_37 & 1813 $\times$ 1813 & 11246 & 0.003421389 & subsequent circuit simulation problem \\
adder\_dcop\_21 & 1813 $\times$ 1813 & 11246 & 0.003421389 & subsequent circuit simulation problem \\
CAG\_mat1916 & 1916 $\times$ 1916 & 195985 & 0.053386546 & combinatorial problem \\
G29 & 2000 $\times$ 2000 & 19990 & 0.009995 & undirected weighted random graph \\
rail2586 & 2586 $\times$ 923269 & 8011362 & 0.003355441 & linear programming problem \\
ex12 & 3973 $\times$ 3973 & 42092 & 0.005333256 & computational fluid dynamics problem \\
ex12 & 3973 $\times$ 3973 & 42092 & 0.002666628 & computational fluid dynamics problem \\
sts4098 & 4098 $\times$ 4098 & 38227 & 0.004552567 & structural problem \\
M80PI\_n & 4182 $\times$ 4182 & 10261 & 0.000586707 & eigenvalue/model reduction problem \\
t2dal\_e & 4257 $\times$ 4257 & 4257 & 0.000234907 & duplicate model reduction problem \\
seymourl & 4944 $\times$ 6316 & 38493 & 0.001232711 & linear programming problem \\
n4c6-b13 & 6300 $\times$ 25605 & 88200 & 0.000546768 & combinatorial problem \\
c-36 & 7479 $\times$ 7479 & 36710 & 0.000656292 & optimization problem \\
bcsstm38 & 8032 $\times$ 8032 & 7842 & 0.000121557 & structural problem \\
nemeth16 & 9506 $\times$ 9506 & 298259 & 0.003300638 & subsequent theoretical/quantum chemistry problem \\
flowmeter5 & 9669 $\times$ 9669 & 67391 & 0.00072084 & model reduction problem \\
sit100 & 10262 $\times$ 10262 & 34094 & 0.000323753 & 2D/3D problem \\
c-42 & 10471 $\times$ 10471 & 60378 & 0.000550684 & optimization problem \\
c-44 & 10728 $\times$ 10728 & 47864 & 0.000831767 & optimization problem \\
ch7-6-b3 & 12600 $\times$ 4200 & 50400 & 0.000952381 & combinatorial problem \\
bayer10 & 13436 $\times$ 13436 & 94926 & 0.00052583 & chemical process simulation problem \\
cyl6 & 13681 $\times$ 13681 & 363961 & 0.0038891 & structural problem \\
ch7-6-b4 & 15120 $\times$ 12600 & 75600 & 0.000396825 & combinatorial problem \\
psse0 & 26722 $\times$ 11028 & 102432 & 0.000347592 & power network problem \\
2D\_27628\_bjtcai & 27628 $\times$ 27628 & 442898 & 0.00058023 & semiconductor device problem \\
c-55 & 32780 $\times$ 32780 & 218115 & 0.000405973 & optimization problem \\
scagr7-2r & 32847 $\times$ 46679 & 120141 & 0.000078356 & linear programming problem sequence \\
stat96v3 & 33841 $\times$ 1113780 & 3317736 & 0.000088023 & linear programming problem \\
jan99jac100sc & 34454 $\times$ 34454 & 215862 & 0.000181843 & economic problem \\
mosfet2 & 46994 $\times$ 46994 & 1499460 & 0.000678969 & semiconductor device problem \\
Andrews & 60000 $\times$ 60000 & 410077 & 0.00011391 & computer graphics/vision problem \\
GaAsH6 & 61349 $\times$ 61349 & 1721579 & 0.000457417 & theoretical/quantum chemistry problem \\
lhr71 & 70304 $\times$ 70304 & 1528092 & 0.000309164 & chemical process simulation problem \\
oilpan & 73752 $\times$ 73752 & 1835470 & 0.000337442 & structural problem \\
t3dh\_a & 79171 $\times$ 79171 & 2215638 & 0.000353481 & duplicate model reduction problem \\
x104 & 108384 $\times$ 108384 & 5138004 & 0.000437385 & structural problem \\
cage12 & 130228 $\times$ 130228 & 2032536 & 0.000119848 & directed weighted graph \\
tp-6 & 142752 $\times$ 1014301 & 11537419 & 0.000079681 & linear programming problem \\
marine1 & 400320 $\times$ 400320 & 6226538 & 0.000038853 & chemical oceanography problem \\
t2em & 921632 $\times$ 921632 & 4590832 & 0.000005405 & electromagnetics problem \\
G3\_circuit & 1585478 $\times$ 1585478 & 4623152 & 0.000001839 & circuit simulation problem \\

\bottomrule
\end{tabular}

\label{tab:appendix-matrix-info}
\end{table*}

\end{document}